\documentclass[10pt,final,journal,letterpaper,twoside,twocolumn]{IEEEtran}
\usepackage{cite}
\usepackage[cmex10]{amsmath}
\usepackage{placeins}
\usepackage{graphicx,epsfig,amssymb}
\usepackage{anysize}
\marginsize{0.8in}{0.8in}{0.25in}{0.25in} 
\usepackage{algorithm} 
\usepackage{algorithmic} 

\begin{document}
\title{Energy Efficiency Optimization for MIMO-OFDM Mobile Multimedia Communication Systems with QoS Constraints}
\author{Xiaohu~Ge,~\IEEEmembership{Senior~Member,~IEEE,}
        Xi Huang,
        Yuming Wang,~\IEEEmembership{Member,~IEEE,}
        Min Chen,~\IEEEmembership{Senior~Member,~IEEE,}
        Qiang Li,~\IEEEmembership{Member,~IEEE,}
        Tao Han,~\IEEEmembership{Member,~IEEE,}
        and Cheng-Xiang Wang,~\IEEEmembership{Senior~Member,~IEEE}
\thanks{\scriptsize{Copyright (c) 2013 IEEE. Personal use of this material is permitted. However, permission to use this material for any other purposes must be obtained from the IEEE by sending a request to pubs-permissions@ieee.org.

X. Ge, X. Huang, Yuming Wang(corresponding author), Qiang Li and Tao Han are with the Department of Electronics and Information Engineering, Huazhong University of Science and Technology, Wuhan 430074, Hubei, P. R. China (email: \{xhge,ymwang,qli\_patrick,hantao\}@mail.hust.edu.cn, hbszhzm@163.com).}}
\thanks{\scriptsize{M. Chen is with the School of Computer Science and Technology,Huazhong University of Science and Technology, Wuhan 430074, Hubei, P. R. China. (email: minchen2012@hust.edu.cn).}}
\thanks{\scriptsize{C.-X. Wang is with the Joint Research Institute for Signal and Image Processing, School of Engineering \&Physical Sciences, Heriot-Watt University, Edinburgh, EH14 4AS, UK. (email: cheng-xiang.wang@hw.ac.uk).}}
\thanks{\scriptsize{The authors would like to acknowledge the support from the National Natural Science Foundation of China (NSFC) under the grants 61103177 and 61271224, NFSC Major International Joint Research Project under the grant 61210002, National 863 High Technology Program of China under the grant 2009AA01Z239 and the Ministry of Science and Technology (MOST) of China under the grants 0903 and 2014DFA11640, the Hubei Provincial Science and Technology Department under the grant 2011BFA004, the Fundamental Research Funds for the Central Universities under the grant 2011QN020 and the Opening Project of the Key Laboratory of Cognitive Radio and Information Processing (Guilin University of Electronic Technology), Ministry of Education (No. 2013KF01). This research is partially supported by EU FP7-PEOPLE-IRSES, project acronym S2EuNet (grant no. 247083), project acronym WiNDOW (grant no. 318992) and project acronym CROWN (grant no. 610524).}}
}
\maketitle

\begin{abstract}
It is widely recognized that besides the quality of service (QoS), the energy efficiency is also a key parameter in designing and evaluating mobile multimedia communication systems, which has catalyzed great interest in recent literature. In this paper, an energy efficiency model is first proposed for multiple-input multiple-output orthogonal-frequency-division-multiplexing (MIMO-OFDM) mobile multimedia communication systems with statistical QoS constraints. Employing the channel matrix singular-value-decomposition (SVD) method, all subchannels are classified by their channel characteristics. Furthermore, the multi-channel joint optimization problem in conventional MIMO-OFDM communication systems is transformed into a multi-target single channel optimization problem by grouping all subchannels. Therefore, a closed-form solution of the energy efficiency optimization is derived for MIMO-OFDM mobile mlutimedia communication systems. As a consequence, an energy-efficiency optimized power allocation (EEOPA) algorithm is proposed to improve the energy efficiency of MIMO-OFDM mobile multimedia communication systems. Simulation comparisons validate that the proposed EEOPA algorithm can guarantee the required QoS with high energy efficiency in MIMO-OFDM mobile multimedia communication systems.
\end{abstract}

\section{Introduction}
\IEEEPARstart{A}{S} the rapid development of the information and communication technology (ICT), the energy consumption problem of ICT industry, which causes about 2\% of worldwide $C{O_2}$ emissions yearly and burdens the electrical bill of network operators \cite{Ge11}, has drawn universal attention. Motivated by the demand for improving the energy efficiency in mobile multimedia communication systems, various resource allocation optimization schemes aiming at enhancing the energy efficiency have become one of the mainstreams in mobile multimedia communication systems, including transmission power allocation \cite{Li11, Raghavendra12}, bandwidth allocation \cite{Liu08, Ding10, Su10}, subchannel allocation \cite{Helonde11}, and etc. Multi-input multi-output (MIMO) technologies can create independent parallel channels to transmit data streams, which improves spectrum efficiency and system capacity without increasing the bandwidth requirement \cite{chengxiang07}. Orthogonal-frequency-division-multiplexing (OFDM) technologies eliminate the multipath effect by transforming frequency selective channels into flat channels. As a combination of MIMO and OFDM technologies, the MIMO-OFDM technologies are widely used in mobile multimedia communication systems. However, how to improve energy efficiency with quality of service (QoS)constraint is an indispensable problem in MIMO-OFDM mobile multimedia communication systems.

The energy efficiency has become one of the hot studies in MIMO wireless communication systems in the last decade \cite{Xiang13, Chen11, Heliot12, Wang13, Hong13, Ku13}. An energy efficiency model for Poisson-Voronoi tessellation
(PVT) cellular networks considering spatial distributions of traffic load and power consumption was proposed \cite{Xiang13}.  The energy-bandwidth efficiency tradeoff in MIMO multihop wireless networks was studied and the effects of different numbers of antennas on the energy-bandwidth efficiency tradeoff were investigated in \cite{Chen11}. An accurate closed-form approximation of the tradeoff between energy efficiency and spectrum efficiency over the MIMO Rayleigh fading channel was derived by considering different types of power consumption model \cite{Heliot12}. A relay cooperation scheme was proposed to investigate the spectral and energy efficiencies tradeoff in multicell MIMO cellular networks \cite{Wang13}.The energy efficiency-spectral efficiency tradeoff of the uplink of a multi-user cellular V-MIMO system with decode-and-forward type protocols was studied in \cite{Hong13}. The tradeoff between spectral and energy efficiency was investigated in the relay-aided multicell MIMO  cellular network by comparing both the signal forwarding and interference forwarding relaying paradigms \cite{Ku13}. In our earlier work, we explored the tradeoff between the operating power and the embodied power contained in the manufacturing process of infrastructure equipments from a life-cycle perspective \cite{Ge11}. In this paper, we further investigate the energy efficiency optimization for  MIMO-OFDM mobile multimedia communication systems.

Based on the Wishart matrix theory \cite{Fisher15, Wishart28, Wishart48, Matthaiou09}, numerous channel models have been proposed in the literature for MIMO communication systems \cite{Zanella08, Jin06, Zanella051, Zanella052, McKay07, Kang03, Kang04, Park06}.
A closed-form joint probability density function (PDF) of eigenvalues of Wishart matrix was derived for evaluating the performance of MIMO communication systems \cite{Zanella08}. Moreover, a closed-form expression for the marginal PDF of the ordered eigenvalues of complex noncentral Wishart matrices was derived to analyze the performance of singular value decomposition (SVD)
in MIMO communication systems with Ricean fading channels \cite{Jin06}. Based on the distribution of eigenvalues of Wishart matrix, the performance of high spectrum efficiency MIMO communication systems with $M$-PSK (Multiple Phase Shift Keying) signals in a flat Rayleigh-fading environment was investigated in terms of symbol error probabilities \cite{Zanella051}. Furthermore, the cumulative density functions (CDF) of the largest and the smallest eigenvalue of a central correlated Wishart matrix were investigated to evaluate the error probability of a MIMO maximal ratio combing (MRC) communication system with perfect channel state information at both transmitter and receiver \cite{Zanella052}. Based on PDF and CDF of the maximum eigenvalue of double-correlated complex Wishart matrices, the exact expressions for the PDF of the output signal-to-noise ratio (SNR) were derived for MIMO-MRC communication systems with Rayleigh fading channels \cite{McKay07}. The closed-form expressions for the outage probability
of MIMO-MRC communication systems with Rician-fading channels were derived under the condition of the largest eigenvalue distribution of central complex Wishart matrices in the noncentral case \cite{Kang03}. Furthermore, The closed-form expressions for the outage probability of MIMO-MRC communication systems with and without co-channel interference were derived by using CDFs of Wishart matrix \cite{Kang04}. Meanwhile, the PDF of the smallest eigenvalue of Wishart matrix was applied to select antennas to improve the capacity of MIMO communication systems \cite{Park06}. However, most existing studies mainly worked on the joint PDF of eigenvalues of Wishart matrix to measure the channel performance for MIMO communication systems. In our study, subchannels' gains derived from the marginal probability distribution of Wishart matrix is investigated to implement energy efficiency optimization in MIMO-OFDM mobile multimedia communication systems.

In conventional mobile multimedia communication systems, many studies have been carried out \cite{Niyato10, Karray10, Wu03, Gursoy09, Tang071, Tang072, Bogucka11}.
In terms of the corresponding QoS demand of different throughput levels in MIMO communication systems, an effective antenna assignment scheme and an access control scheme were proposed in \cite{Niyato10}. A downlink QoS evaluation scheme was proposed from the viewpoint of mobile users in orthogonal frequency-division multiple-access (OFDMA) wireless cellular networks \cite{Karray10}. To guarantee the QoS in wireless networks, a statistical QoS constraint model was built to analyze the queue characteristics of data transmissions \cite{Wu03}. The energy efficiency in fading channels under QoS constraints was analyzed in \cite{Gursoy09}, where the effective capacity was considered as a measure of the maximum throughput under certain statistical QoS constraints. Based on the effective capacity of the block fading channel model, a QoS driven power and rate adaptation scheme over wireless links was proposed for mobile wireless networks \cite{Tang071}. Furthermore, by integrating information theory with the effective capacity, some QoS-driven power and rate adaptation schemes was proposed for diversity and multiplexing systems \cite{Tang072}. Simulation results showed that multi-channel communication systems can achieve both high throughput and stringent QoS at the same time. Aiming at optimizing the energy consumption, the key tradeoffs between energy efficiency and link-level QoS metrics were analyzed in different wireless communication scenarios \cite{Bogucka11}. However, there has been few research work addressing the problem of optimizing the energy efficiency under different QoS constraints in MIMO-OFDM mobile multimedia communication systems.


Motivated by aforementioned gaps, this paper is devoted to the energy efficiency optimization with statistical QoS constraints in MIMO-OFDM mobile multimedia communication systems with statistical QoS constraints which uses a statistical exponent to measure the queue characteristics of data transmission in wireless systems.". All subchannels in MIMO-OFDM communication systems are first grouped by their channel gains. On this basis, a novel subchannel grouping scheme is developed to allocate the corresponding transmission power to each of subchannels in different groups, which simplifies the multi-channel optimization problem to a multi-target single channel optimization problem. The main contributions of this paper are summarized as follows.

\begin{enumerate}
\item An energy efficiency model with statistical QoS constraints is proposed for MIMO-OFDM mobile multi-media communication systems.
\item A subchannel grouping scheme is designed by using the channel matrix single-value-decomposition (SVD) method, which simplifies the multi-channel optimization problem to a multi-target single channel optimization problem. Based on marginal probability density functions (MPDFs) of subchannels in different groups, a closed-form solution of energy efficiency optimization is derived for MIMO-OFDM mobile multimedia communication systems.
\item A novel algorithm is developed to optimize the energy efficiency in MIMO-OFDM mobile multimedia communication systems. Numerical results validate that the proposed algorithm improves the energy efficiency of MIMO-OFDM mobile multimedia communication systems with statistical QoS constraints.
\end{enumerate}
\begin{figure*}[!t]
\vspace{0.1in}
\centerline{\includegraphics[width=14cm,draft=false]{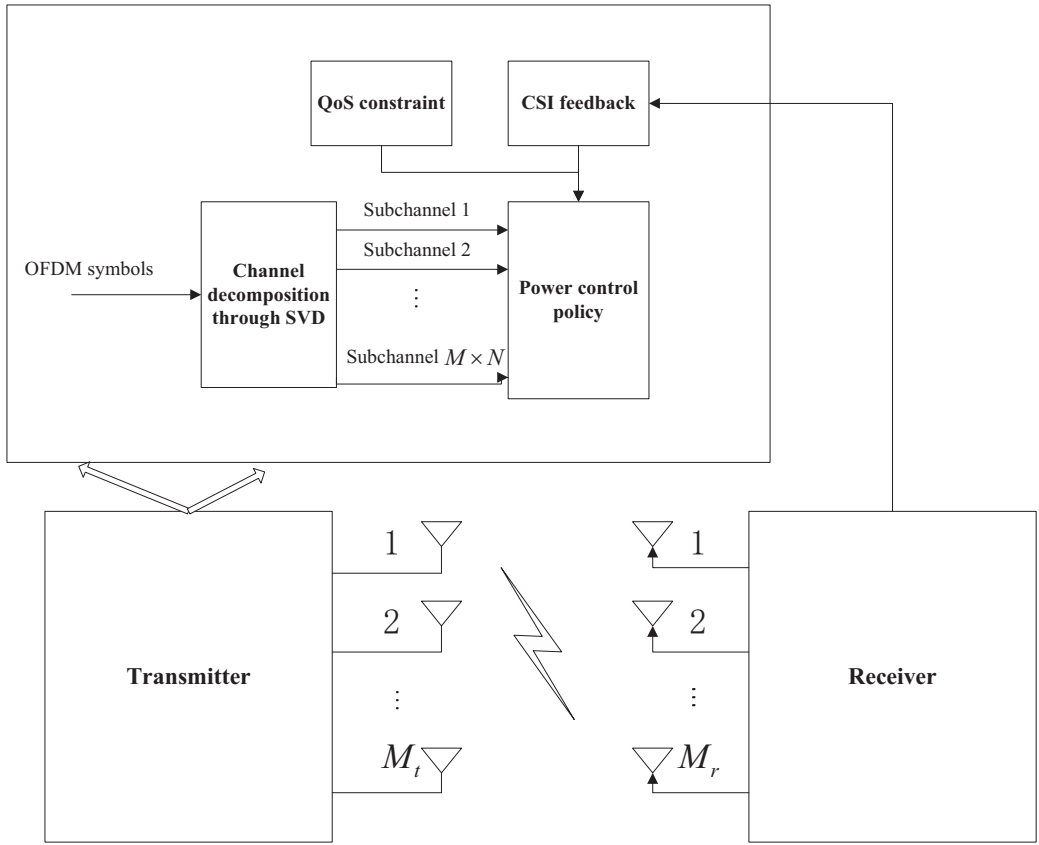}}
\caption{\small MIMO-OFDM system model.}
\label{fig1}
\end{figure*}

The remainder of this paper is organized as follows. The system model is introduced in Section~\ref{sec2}. In Section~\ref{sec3}, the energy efficiency model of MIMO-OFDM mobile multimedia communication systems with statistical QoS constraints is proposed. Based on the subchannel grouping scheme, a closed-form solution of energy efficiency optimization is derived for MIMO-OFDM mobile multimedia communication systems in Section~\ref{sec4}. Moreover, a novel transmission power allocation algorithm is presented. Numerical results are illustrated in Section~\ref{sec5}. Finally, Section~\ref{sec6} concludes the paper.




\section{System Model}
\label{sec2}
The MIMO-OFDM mobile multimedia communication system is illustrated in Fig.~\ref{fig1}. It has a ${M_r} \times {M_t}$ antenna matrix, $N$ subcarriers and $S$ OFDM symbols, where ${M_t}$ is the number of transmit antennas and ${M_r}$ is the number of receive antennas. We denote $B$ as the system bandwidth and ${T_f}$ as the frame duration. The OFDM signals are assumed to be transmitted within a frame duration. Then the received signal of MIMO-OFDM communication system can be expressed as follows:
\begin{equation}\label{eq1}
{{\bf{y}}_k}[i] = {{\bf{H}}_k}{{\bf{x}}_k}[i] + {\bf{n}},
\tag{1}
\end{equation}
where ${{\bf{y}}_k}[i]$ and ${{\bf{x}}_k}[i]$ are the received signal vector and transmitted signal vector at the $k$th $(k = 1,2,...,N)$ subcarrier of the $i$th $\left(i = 1,2,...,S\right)$ OFDM symbol, respectively. ${{\mathbf{H}}_k}$ is the frequency-domain channel matrix at the $k$th subcarrier and ${\mathbf{n}}$ is the additive noise vector. Let $\mathbb{C}$ denote the complex space, then we have ${{\mathbf{y}}_k} \in {\mathbb{C}^{{M_r}}}$, ${{\mathbf{x}}_k} \in {\mathbb{C}^{{M_t}}}$, ${{\mathbf{H}}_k} \in {\mathbb{C}^{{M_r} \times {M_t}}}$, and ${\mathbf{n}} \in {\mathbb{C}^{{M_r}}}$. Without loss of generality, we assume ${\mathbf{{\rm E}}}\{ {\mathbf{n}}{{\mathbf{n}}^H}\}  = {{\mathbf{I}}^{{M_r} \times {M_r}}}$, where ${\mathbf{{\rm E}}}\{  \cdot \} $ denotes the expectation operator.

Discrete-time channels are assumed to experience a block-fading, in which the frame duration is shorter than the channel coherence time. Based on this assumption, the channel gain is invariant within a frame duration ${T_f}$, but varies independently from one frame to another. In each frame duration, the channel at each subcarrier is divided into $M$ $\left(M = \min ({M_t},{M_r})\right)$ parallel SISO channels by the SVD method. As a consequence, a total number of $M \times N$ parallel space-frequency subchannels can be generated in each OFDM symbol. Transmitters are assumed to obtain the channel state information (CSI) from receivers without delay via feedback channels. Furthermore, an average transmission power constraint $\overline {P}$ is configured for each subchannel in the MIMO-OFDM communication system. With this average transmission power constraint, transmitters are able to perform power control adaptively according to the feedback CSI and system QoS constraints, so that the energy efficiency in the MIMO-OFDM mobile multimedia communication system can be optimized. To facilitate reading, the notations and symbols used in this paper are listed in TABLE~\ref{tab1}.
\begin{table*}
\centering
\caption{NOTATIONS AND SYMBOLS USED IN THE PAPER}
\begin{tabular}{l|l}
\hline Symbol & Definition/explanation \\
\hline
${M_t}$ & The number of transmit antennas \\
${M_r}$ & The number of receive antennas \\
$M$ & $M = \min ({M_t},{M_r})$ \\
$N$ & The number of subcarriers \\
$S$ & The number of OFDM symbols \\
$B$ & The system bandwidth \\
${T_f}$ & The frame duration \\
${{\mathbf{y}}_k}$ & The received signal vector at the $k$th subcarrier of the $i$th OFDM symbol \\
${{\mathbf{x}}_k}$ & The transmitted signal vector at the $k$th subcarrier of the $i$th OFDM symbol \\
${{\mathbf{H}}_k}$ & The frequency-domain channel matrix at the $k$th subcarrier \\
${\mathbf{n}}$ & The additive noise vector \\
$M$ & The number of parallel SISO channels at each subcarrier\\
$\overline {P}$ & The average transmission power constraint for a subchannel \\
$\lambda$ & The subchannel gain \\
${\lambda _{m,k}}$ & The channel gain of the $m$th subchannel at the $k$th subcarrier \\
$\Lambda $ & The transmission power allocation threshold over a subchannel \\
${\Lambda _{m,k}}$ & The transmission power allocation threshold of the $m$th subchannel at the $k$th subcarrier \\
${\Lambda _n}$ & The transmission power allocation threshold of the $n$th grouped subchannels \\
$\eta $ & The energy efficiency of MIMO-OFDM mobile multimedia communication systems \\
${\eta _{\text{opt}}}$ & The optimized energy efficiency  \\
$\theta $ & The QoS statistical exponent \\
$\beta $ & The normalized QoS exponent \\
${C_{total}}(\theta )$ & The total effective capacity \\
${C_e}{(\theta )_{m,k}}$ & The effective capacity of the $m$th subchannel over the $k$th subcarrier \\
${C_e}(\theta )$ & The effective capacity for a subchannel with QoS constraint \\
${C_e}{(\theta )_{{\text{opt\_}}\operatorname{n} }}$ & The optimized effective capacity of the $n$th grouped subchannels \\
${\mathbf{E}}\left\{ {{P_{total}}} \right\}$ & The expectation of the total transmission power \\
$R$ & The instantaneous bit rate within a frame duration \\
$\mu (\theta ,\lambda )$ & The transmission power allocated over a subchannel \\
${\mu _{m,k}}(\theta ,\lambda )$ & The transmission power allocated over the $m$th subchannel at the $k$th orthogonal subcarrier \\
${\mu _{{\text{opt}}}}(\theta ,\lambda )$ & The optimized transmission power allocated over a channel \\
${\mu _{{\text{opt\_n}}}}(\theta ,\lambda )$ & The optimized transmission power allocated for subchannels in the $n$th group \\
${p_{{\Gamma _{m,k}}}}(\lambda )$ & The channel gain MPDF of the $m$th subchannel at the $k$th orthogonal subcarrier \\
${p_{{\Gamma _n}}}(\lambda )$ & The channel gain MPDF of the subchannels over the $n$th grouped subchannels \\
$p({\lambda _1},{\lambda _2},...,{\lambda _M})$ & The joint PDF of ordered eigenvalues of a Wishart matrix \\
${K_{M,Q}}$ & The normalizing factor \\
\hline
\end{tabular}
\label{tab1}
\end{table*}

\section{Energy Efficiency Modeling of MIMO-OFDM Mobile Multimedia Communication Systems}
\label{sec3}
Applying the SVD method to the channel matrix ${{\mathbf{H}}_k}$ at each subcarrier, where ${{\mathbf{H}}_k} \in {\mathbb{C}^{{M_r} \times {M_t}}}$ $\left(k = 1,2,...,N\right)$, we have
\begin{equation}\label{eq2}
  {{\mathbf{H}}_k} = {{\mathbf{U}}_k}\sqrt {{{\mathop {\mathbf{\widetilde{\Delta}}}}_k}} {\mathbf{V}}_k^H,
\tag{2}
\end{equation}
where ${{\mathbf{U}}_k} \in {\mathbb{C}^{{M_r} \times {M_r}}}$ and ${{\mathbf{V}}_k} \in {\mathbb{C}^{{M_t} \times {M_t}}}$ are unitary matrices. When ${M_r} \geqslant {M_t}$, we have block matrix ${\mathop {\mathbf{\widetilde{\Delta}}}_k} = [{{\mathbf{\Delta }}_k},{{\mathbf{0}}_{{M_r},{M_t} - {M_r}}}]$; otherwise when ${M_r} < {M_t}$ ��we have ${\mathop {\mathbf{\widetilde{\Delta}}}_k} = {[{{\mathbf{\Delta }}_k},{{\mathbf{0}}_{{M_t},{M_r} - {M_t}}}]^T}$, where ${{\mathbf{\Delta }}_k} = diag({\lambda _{1,k}},...,{\lambda _{M,k}})$ and ${\lambda _{m,k}} \geqslant 0,\forall m = 1,...,M,k = 1,...,N$. $\{ {\lambda _{m,k}}\} _{m = 1}^M$ denotes the subchannel gain set at the $k$th subcarrier.. In this way, the MIMO channel at each subcarrier is decomposed into $M$ parallel SISO subchannels by SVD method. Therefore, $M \times N$ parallel space-frequency subchannels are obtained at $N$ orthogonal subcarriers for each OFDM symbol.

In traditional energy efficiency optimization researches, Shannon capacity is usually used as the index which measures the system output. However, in any practical wireless communication systems, the system capacity is obviously less than Shannon capacity, especially in the scenario with strict QoS constraint. In this paper, the effective capacity of each subchannel is taken as the practical data rate with certain QoS constraint.The total effective capacity of $M \times N$ subchannels is configured as the system output and the total transmission power allocated to $M \times N$ subchannels is configured as the system input. As a consequence, the energy efficiency of MIMO-OFDM mobile multimedia communication systems is defined as follows
\begin{equation}\label{eq3}
  \eta  = \frac{{{C_{total}}(\theta )}}
{{{\mathbf{E}}\left\{ {{P_{total}}} \right\}}}\;\; = \frac{{\sum\limits_{m = 1}^M {\sum\limits_{k = 1}^N {{C_e}{{(\theta )}_{m,k}}} } }}
{{{\mathbf{E}}\left\{ {{P_{total}}} \right\}}},
\tag{3}
\end{equation}
where ${C_e}{(\theta )_{m,k}} (m = 1,2,...,M,{\kern 1pt} {\kern 1pt} k = 1,2,...,N)$  is the effective capacity of the $m$th subchannel over the $k$th subcarrier, and ${\mathbf{E}}\left\{ {{P_{total}}} \right\}$ is the expectation of the total transmission power allocated to all $M \times N$ subchannels. $\theta $ is the QoS statistical exponent, which indicates the exponential decay rate of QoS violation probabilities \cite{Tang071}. A smaller $\theta $ corresponds to a slower decay rate, which implies that the multimedia communication system provides a looser QoS guarantee; while a larger $\theta $ leads to a faster decay rate, which means that a higher QoS requirement should be supported.

Practical MIMO-OFDM mobile multimedia communication systems involve multiple services, such as speech and video services, which are sensitive to the delay parameter. Different services in MIMO-OFDM mobile multimedia communication systems have different QoS constraints. In view of this, the effective capacity of each subchannel depends on the corresponding QoS constraint. A statistical QoS constraint is adopted to evaluate the effective capacity of each subchannel which is calculated as the system practical output in MIMO-OFDM mobile multimedia communication systems. Assuming the fading process over wireless channels is independent among frames and keeps invariant within a frame duration, the effective capacity ${C_e}(\theta )$ for a subchannel with QoS statistical exponent $\theta $ in MIMO-OFDM mobile multimedia communication systems is expressed as follows \cite{Tang071}
\begin{equation}\label{eq4a}
  {C_e}(\theta ) =  - \frac{1}
{\theta }\log \left( {{\mathbf{{\rm E}}}\left\{ {{e^{ - \theta R}}} \right\}} \right),
\tag{4a}
\end{equation}
\begin{equation}\label{eq4b}
  R = {T_f}B{\log _2}(1 + \mu (\theta ,\lambda )\lambda ),
\tag{4b}
\end{equation}
where $R$ denotes the instantaneous bit rate within a frame duration, $\lambda$ denotes the subchannel gain, and $\mu (\theta ,\lambda )$ denotes the transmission power allocated to a subchannel.

After SVD of channel matrices at $N$ orthogonal subcarriers, $M \times N$ parallel subchannels are obtained. The channel gain over each of these $M \times N$ parallel subchannels follows a marginal probability distribution (MPDF). Assuming ${p_{{\Gamma _{m,k}}}}(\lambda )$ as the MPDF of channel gain over the $m$th $\left(m = 1,2,...,M\right)$ subchannel at the $k$th $\left(k = 1,2,...,N\right)$ orthogonal subcarrier, then the corresponding effective capacity ${C_e}{(\theta )_{m,k}}$ over the $m$th subchannel at the $k$th orthogonal subcarrier is derived as (5).
\begin{figure*}[!t]
\begin{equation}\label{eq5}
  {C_e}{(\theta )_{m,k}} =  - \frac{1}
{\theta }\log \left( {\int_0^\infty  {{e^{ - \theta {T_f}B{{\log }_2}(1 + {\mu _{m,k}}(\theta ,\lambda )\lambda )}}{p_{{\Gamma _{m,k}}}}(\lambda )d\lambda } } \right),
\tag{5}
\end{equation}
\end{figure*}
where ${\mu _{m,k}}(\theta ,\lambda )$ is the transmission power allocated to the $m$th subchannel at the $k$th orthogonal subcarrier.

Considering the practical power consumption limitation at transmitters, an average transmission power constraint ${\overline{P}}$ over each subchannel is derived as (6).
\begin{figure*}[!t]
\begin{equation}\label{eq6}
  \overline{P} = \int_0^\infty  {{\mu _{m,k}}(\theta ,\lambda ){p_{{\Gamma _{m,k}}}}(\lambda )d\lambda } {\kern 1pt} {\kern 1pt} {\kern 1pt} \;\;(\forall m = 1,2,...,M,{\kern 1pt} {\kern 1pt} k = 1,2,...,N).
\tag{6}
\end{equation}
\end{figure*}
With the average transmission power constraint, the expectation of transmission power ${\mathbf{E}}\left\{ {{P_{total}}} \right\}$ is given by
\begin{equation}\label{eq7}
{\mathbf{E}}\left\{ {{P_{total}}} \right\} = \overline{P} \times M \times N.
\tag{7}
\end{equation}

Substituting expression (\ref{eq6}) and (\ref{eq7}) into (\ref{eq3}), we derive the energy efficiency model as (8).
\begin{figure*}[!t]
\begin{equation}\label{eq8}
  \begin{array}{*{20}{c}}
   \eta  \hfill  \\
 \end{array}  = \frac{{\sum\limits_{m = 1}^M {\sum\limits_{k = 1}^N { - \frac{1}
{\theta }\log \left( {\int_0^\infty  {{e^{ - \theta {T_f}B{{\log }_2}(1 + {\mu _{m,k}}(\theta ,\lambda )\lambda )}}{p_{{\Gamma _{m,k}}}}(\lambda )d\lambda } } \right)} } }}
{{\overline{P} \times M \times N}}.
\tag{8}
\end{equation}
\end{figure*}

From (\ref{eq8}), the energy efficiency of MIMO-OFDM mobile multimedia communication systems depends on the MPDF ${p_{{\Gamma _{m,k}}}}(\lambda ) \left(m = 1,2,...,M,{\kern 1pt} {\kern 1pt} k = 1,2,...,N\right)$ over $M \times N$ subchannels. Since there is a relationship between the MPDF ${p_{{\Gamma _{m,k}}}}(\lambda )$ and statistical characteristics of the subchannel, the marginal distribution characteristics of each subchannel gain is investigated to optimize the energy efficiency in MIMO-OFDM mobile multimedia communication systems.

\section{Energy Efficiency Optimization of Mobile Multimedia Communication Systems}
\label{sec4}
In MIMO wireless communication systems, statistical characteristics of channel gain depend on the eigenvalues' distribution of Hermitian channel matrix ${\mathbf{H}}{{\mathbf{H}}^H}$, where ${\mathbf{H}}$ is the channel matrix \cite{Chiani03, Telatar99, Kang06}. When the elements of ${\mathbf{H}}$ are complex valued with real and imaginary parts each governed by a normal distribution $\mathfrak{N}(0,1/2)$ with mean value 0 and variance value 1/2, the Hermitian channel matrix ${\mathbf{W}} = {\mathbf{H}}{{\mathbf{H}}^H}$ is called a central Wishart channel matrix \cite{Fisher15, Wishart28, Wishart48, Zanella08}. In this case, ${\mathbf{E}}\left\{ {\mathbf{H}} \right\} = {\mathbf{0}}$ and wireless channels have the Rayleigh fading characteristic. If ${\mathbf{E}}\left\{ {\mathbf{H}} \right\} \ne {\mathbf{0}}$, ${\mathbf{W}} = {\mathbf{H}}{{\mathbf{H}}^H}$ is a noncentral Wishart channel matrix and wireless channels have the Rician fading characteristic \cite{Jin06}.

Based on SVD results of wireless channel matrix, subchannels at each orthogonal subcarrier are sorted in a descending order of channel gains. Starting from the joint PDF of eigenvalues of Wishart channel matrix, the channel gain MPDF of subchannels ordered at the $m$th position in the descending order of channel gains is derived. Furthermore, all subchannels at $N$  subcarriers are grouped according to their MPDFs. In terms of subchannel grouping results, a closed-form solution is derived to optimize the energy efficiency of MIMO-OFDM mobile multimedia communication systems in this section.

\subsection{Optimization Solution of Energy Efficiency}

\label{sec4-1}
To maximize the energy efficiency of MIMO-OFDM mobile multimedia communication systems with statistical QoS constraints, the optimization problem can be formulated as (9).
\begin{figure*}[!t]
\begin{equation}\label{eq9}
\begin{split}
{\eta _{\text{opt}}} &= \max \left\{ {\frac{{\sum\limits_{m = 1}^M {\sum\limits_{k = 1}^N { - \frac{1}
{\theta }\log \left( {\int_0^\infty  {{e^{ - \theta {T_f}B{{\log }_2}(1 + {\mu _{m,k}}(\theta ,\lambda )\lambda )}}{p_{{\Gamma _{m,k}}}}(\lambda )d\lambda } } \right)} } }}{{\overline {P} \times M \times N}}} \right\} \\
              &= \frac{{\max \left\{ {\sum\limits_{m = 1}^M {\sum\limits_{k = 1}^N { - \frac{1}{\theta }\log \left( {\int_0^\infty  {{e^{ - \theta {T_f}B{{\log }_2}(1 + {\mu _{m,k}}(\theta ,\lambda )\lambda )}}{p_{{\Gamma _{m,k}}}}(\lambda )d\lambda } } \right)} } } \right\}}}{{\overline {P} \times M \times N}},\\
s.t.:\\
\end{split}
\tag{9}
\end{equation}
\end{figure*}

\begin{figure*}[!t]
\begin{equation}\label{eq10}
  \int_0^\infty  {{\mu _{m,k}}(\theta ,\lambda )}{p_{{\Gamma _{m,k}}}}(\lambda )d\lambda  \le \overline {P},\forall m = 1,2,...,M,{\kern 1pt} {\kern 1pt} k = 1,2,...,N.
\tag{10}
\end{equation}
\end{figure*}
where ${\eta _{\text{opt}}}$ is the optimized energy efficiency.

From the problem formulation in (\ref{eq9}) and (\ref{eq10}), it is remarkable that the energy efficiency of MIMO-OFDM mobile multimedia communication systems depends on transmission power allocation results ${\mu _{m,k}}(\theta ,\lambda )$ over $M \times N$ subchannels. In this case, the optimization problem in (\ref{eq9}) and (\ref{eq10}) is a multi-channel optimization problem, which is intractable to obtain a closed-form solution in mathematics.

In most studies on MIMO wireless communication systems, the energy efficiency optimization problem is solved by a single channel optimization model \cite{Tang072}. How to change the multi-channel energy efficiency optimization problem into the single channel energy efficiency optimization problem and derive a closed-form solution are great challenges in this paper. Without loss of generality, the optimized transmission power allocation of single subchannel ${\mu _{{\text{opt}}}}(\theta ,\lambda )$ is expressed as follows \cite{Tang072}
\begin{equation}\label{eq11a}
 {\mu _{{\text{opt}}}}(\theta ,\lambda ) = \left\{ {\begin{array}{*{20}{c}}
   {\frac{1}
{{{\Lambda ^{\frac{1}
{{\beta  + 1}}}}{\lambda ^{\frac{\beta }
{{\beta  + 1}}}}}} - \frac{1}
{\lambda },{\kern 1pt} {\kern 1pt} {\kern 1pt} {\kern 1pt} {\kern 1pt} {\kern 1pt} {\kern 1pt} {\kern 1pt} {\kern 1pt} {\kern 1pt} \lambda  \geqslant \Lambda }  \\
   {{\kern 1pt} {\kern 1pt} {\kern 1pt} {\kern 1pt} {\kern 1pt} {\kern 1pt} {\kern 1pt} {\kern 1pt} {\kern 1pt} {\kern 1pt} {\kern 1pt} {\kern 1pt} {\kern 1pt} {\kern 1pt} {\kern 1pt} {\kern 1pt} {\kern 1pt} {\kern 1pt} {\kern 1pt} {\kern 1pt} {\kern 1pt} {\kern 1pt} {\kern 1pt} {\kern 1pt} {\kern 1pt} {\kern 1pt} {\kern 1pt} {\kern 1pt} {\kern 1pt} {\kern 1pt} {\kern 1pt} 0,{\kern 1pt} {\kern 1pt} {\kern 1pt} {\kern 1pt} {\kern 1pt} {\kern 1pt} {\kern 1pt} {\kern 1pt} {\kern 1pt} {\kern 1pt} {\kern 1pt} {\kern 1pt} {\kern 1pt} {\kern 1pt} {\kern 1pt} {\kern 1pt} {\kern 1pt} {\kern 1pt} {\kern 1pt} {\kern 1pt} {\kern 1pt} {\kern 1pt} {\kern 1pt} {\kern 1pt} {\kern 1pt} {\kern 1pt} {\kern 1pt} {\kern 1pt} {\kern 1pt} {\kern 1pt} {\kern 1pt} {\kern 1pt} {\kern 1pt} {\kern 1pt} {\kern 1pt} {\kern 1pt} {\kern 1pt} {\kern 1pt} \lambda  < \Lambda }  \\

 \end{array} } \right.,
\tag{11a}
\end{equation}
\begin{equation}\label{eq11b}
  \beta  = \theta {T_f}B/\log 2,
\tag{11b}
\end{equation}
where $\Lambda $ is the transmission power allocation threshold over a subchannel and $\beta$ is the normalized QoS exponent.

It is critical to determine the transmission power allocation threshold $\Lambda $ for the implemention of optimized transmission power allocation in (\ref{eq11a}). An average transmission power constraint $\overline {P}$ is configured for each subchannel, thus the transmission power allocation threshold of each subchannel should satisfy the following constraint
\begin{equation}\label{eq12}
  \int_{{\Lambda _{m,k}}}^\infty  {\left( {\frac{1}
{{\Lambda _{m,k}^{\frac{1}
{{\beta  + 1}}}{\lambda ^{\frac{\beta }
{{\beta  + 1}}}}}} - \frac{1}
{\lambda }} \right){p_{{\Gamma _{m,k}}}}(\lambda )d\lambda  \le \overline {P}},
\tag{12}
\end{equation}
where ${\Lambda _{m,k} \left(m = 1,2,...,M,{\kern 1pt} {\kern 1pt} k = 1,2,...,N\right)}$ is the transmission power allocation threshold of the $m$th subchannel at the $k$th subcarrier.

Assuming that the channel matrix ${{\mathbf{H}}_k} \left(k = 1,2,...,N\right)$ at each subcarrier is a complex matrix and its elements are complex valued with real and imaginary parts each governed by a normal distribution $\mathfrak{N}(0,1/2)$ with mean value 0 and variance value 1/2, then elements of ${{\mathbf{H}}_k}$ follow an independent and identically distributed (i.i.d.) circular symmetric complex Gaussian distribution with zero-mean and unit-variance. In this case, wireless channels between transmit and receive antennas are Reyleign fading channels with unit energy.

Denote $Q = \max ({M_t},{M_r})$, and set $\widetilde {\mathbf{W}}$ as a $M \times M$ Hermitian matrix:
\begin{equation}\label{eq13}
  \widetilde {\mathbf{W}} = \left\{ {\begin{array}{*{20}{c}}
   {{{\mathbf{H}}_k}{\mathbf{H}}_k^H}  \\
   {{\mathbf{H}}_k^H{{\mathbf{H}}_k}}  \\

 \end{array} } \right.{\kern 1pt} {\kern 1pt} {\kern 1pt} {\kern 1pt} {\kern 1pt} {\kern 1pt} {\kern 1pt} {\kern 1pt} {\kern 1pt} {\kern 1pt} {\kern 1pt} {\kern 1pt} {\kern 1pt} {\kern 1pt} {\kern 1pt} {\kern 1pt} {\kern 1pt} {\kern 1pt} {\kern 1pt} {\kern 1pt} {\kern 1pt} {\kern 1pt} {\kern 1pt} \begin{array}{*{20}{c}}
   {{M_r} < {M_t}}  \\
   {{M_r} \geqslant {M_t}}  \\

 \end{array} {\kern 1pt} ,
\tag{13}
\end{equation}
then $\widetilde {\mathbf{W}}$ is a central Wishart matrix. The joint PDF of ordered eigenvalues of $\widetilde {\mathbf{W}}$ follows Wishart distributions \cite{Edelman89} as (14).
\begin{figure*}[!t]
\begin{equation}\label{eq14}
  p({\lambda _1},{\lambda _2},...,{\lambda _M}) = K_{M,Q}^{ - 1}{e^{ - \sum\limits_{i = 1}^M {{\lambda _i}} }}\prod\limits_{i = 1}^M {\lambda _i^{Q - M}} \prod\limits_{1 \leqslant i \leqslant j \leqslant M} {{{({\lambda _i} - {\lambda _j})}^2}} ,
\tag{14}
\end{equation}
\end{figure*}
where ${\lambda _1},{\lambda _2},...,{\lambda _{_M}} \left({\lambda _1} \geqslant {\lambda _2} \geqslant ... \geqslant {\lambda _{_M}}\right)$ are ordered eigenvalues of $\widetilde {\mathbf{W}}$, ${K_{M,Q}}$ is a normalizing factor which is denoted as follows:
\begin{equation}\label{eq15}
  {K_{M,Q}} = \prod\limits_{i = 1}^M ( Q - i)!(M - i)! .
\tag{15}
\end{equation}

Based on SVD results of channel matrix ${{\mathbf{H}}_k}$, ordered eigenvalues of matrix ${\mathbf{H}}_k^H{{\mathbf{H}}_k}$ are denoted by elements ${\lambda _{1,k}},{\lambda _{2,k}}...,{\lambda _{M,k}}$ of diagonal matrix ${{\mathbf{\Delta }}_k}$. That means subchannel gains ${\lambda _{1,k}},...,{\lambda _{M,k}}$ at the $k$th subcarrier can be denoted by eigenvalues of the Wishart matrix $\widetilde {\mathbf{W}}$. When subchannel gains at each subcarrier are sorted in a descending order, i.e., $\forall 1 \leqslant i \leqslant j \leqslant M,\;\;1 \leqslant k \leqslant N,\;{\lambda _{i,k}} \geqslant {\lambda _{j,k}}$, the ordered subchannel gains can be denoted by the ordered eigenvalues ${\lambda _1},{\lambda _2},...,{\lambda _{_M}} \left({\lambda _1} \geqslant {\lambda _2} \geqslant ... \geqslant {\lambda _{_M}}\right)$ of Wishart matrix $\widetilde {\mathbf{W}}$, which follow the joint PDF $p({\lambda _1},{\lambda _2},...,{\lambda _M})$ of the ordered eigenvalues of Wishart matrix $\widetilde {\mathbf{W}}$. After subchannel gains at each subcarrier are sorted in a descending order, the MPDF of the $m$th ($1 \leqslant n \leqslant M$) subchannel gain at the $k$th subcarrier ${p_{{\Gamma _{m,k}}}}(\lambda )$ is derived as (16).
\begin{figure*}[!t]
\begin{equation}\label{eq16}
  {p_{{\Gamma _{m,k}}}}(\lambda ) = \underbrace {\int {...} }_{M - 1}\int {p({\lambda _1},{\lambda _2},...,{\lambda _M})d{\lambda _i}d{\lambda _{i + 1}}...d{\lambda _j}{\kern 1pt} {\kern 1pt} {\kern 1pt} {\kern 1pt} {\kern 1pt} {\kern 1pt} {\kern 1pt} {\kern 1pt} {\kern 1pt} (1 \leqslant i < j \leqslant M{\kern 1pt} {\kern 1pt} {\kern 1pt} {\text{and}}{\kern 1pt} {\kern 1pt} i \ne n,j \ne n)}.
\tag{16}
\end{equation}
\end{figure*}
\begin{figure*}[!t]
\begin{equation}\label{eq17}
  \begin{gathered}
  \int_{{\Lambda _n}}^\infty  {\left( {\frac{1}
{{\Lambda _n^{\frac{1}
{{\beta  + 1}}}{\lambda ^{\frac{\beta }
{{\beta  + 1}}}}}} - \frac{1}
{\lambda }} \right)\left( {\underbrace {\int {...} }_{M - 1}\int {p({\lambda _1},{\lambda _2},...,{\lambda _M})d{\lambda _i}d{\lambda _{i + 1}}...d{\lambda _j}} } \right)d\lambda  \le \mathop P\limits^ -  ,} {\kern 1pt} {\kern 1pt} {\kern 1pt} {\kern 1pt} {\kern 1pt} {\kern 1pt} {\kern 1pt} {\kern 1pt} {\kern 1pt}  \hfill \\
  {\kern 1pt} (1 \leqslant i < j \leqslant M{\kern 1pt} {\kern 1pt} {\kern 1pt} {\text{and}}{\kern 1pt} {\kern 1pt} i \ne n,j \ne n) \hfill \\
\end{gathered}
\tag{17}
\end{equation}
\end{figure*}
\begin{figure*}[!t]
\begin{equation}\label{eq19}
  {\eta _{{\text{opt}}}} = \frac{{\sum\limits_{n = 1}^M { - \frac{N}
{\theta }\log \left( {\int_0^\infty  {{e^{ - \theta {T_f}B{{\log }_2}(1 + {\mu _{\operatorname{opt} \_n}}(\theta ,\lambda )\lambda )}}{p_{{\Gamma _n}}}(\lambda )d\lambda } } \right)} }}
{{\overline {P} \times M \times N}}
\tag{19}
\end{equation}
\end{figure*}
\begin{figure*}[!t]
\begin{equation}\label{eq20}
  \qquad \qquad \qquad \qquad \quad \: \: =  - \frac{1}
{{\theta \times \overline {P} \times M}}\sum\limits_{n = 1}^M {\log \left( {\int_0^\infty  {{e^{ - \theta {T_f}B{{\log }_2}(1 + {\mu _{\operatorname{opt} \_n}}(\theta ,\lambda )\lambda )}}{p_{{\Gamma _n}}}(\lambda )d\lambda } } \right)}.
\tag{20}
\end{equation}
\end{figure*}
After subchannels at each subcarrier are sorted by subchannel gains, subchannels with the same order position at different orthogonal subcarriers have the identical MPDF based on (\ref{eq16}). According to this property, a subchannel grouping scheme is proposed for subchannels at different orthogonal subcarriers:
\begin{enumerate}
  \item Sort subchannels at each orthogonal subcarriers by a descending order of subchannel gains: ${\lambda _{1,k}} \geqslant {\lambda _{2,k}} \geqslant ... \geqslant {\lambda _{M,k}} \geqslant 0$, $k = 1,2,...,N$.
  \item For $n = 1:M$, select the subchannels with the same order position at different orthogonal subcarriers (${\lambda _{n,1}},{\lambda _{n,2}},...,{\lambda _{n,N}}$) into different channel groups.
  \item Repeat steps 1) and 2) for all OFDM symbols.
  \item $M$ groups with the same order position subchannels are obtained.
\end{enumerate}

Since subchannels in the same group have an identical MPDF, the MPDF of subchannels in the $n$th group ${p_{{\Gamma _{n,k}}}}(\lambda )$ ($1 \leqslant n \leqslant M,1 \leqslant k \leqslant N$) is simply denoted as ${p_{{\Gamma _n}}}(\lambda )$ ($1 \leqslant n \leqslant M$).

Based on the proposed subchannel grouping scheme, we can optimize the effective capacity of each grouped subchannels according to their MPDFs in (\ref{eq16}), in which all subchannels in the same group have an identical MPDF. In this process, the multi-channel joint optimization problem is transformed into a multi-target single channel optimization problem, which significantly reduces the complexity of energy efficiency optimization. Substituting (\ref{eq16}) into (\ref{eq12}), the average power constraint is derived as (17).
where ${\Lambda _n}$ ($1 \leqslant n \leqslant M$) is the transmission power allocation threshold of the $n$th grouped subchannels. Based on the transmission power allocation threshold for each grouped subchannels in (\ref{eq17}), the optimized transmission power allocation for the $n$th grouped subchannels is formulated as follows
\begin{equation}\label{eq18}
  {\mu _{{\text{opt\_n}}}}(\theta ,\lambda ) = \left\{ {\begin{array}{*{20}{c}}
   {\frac{1}
{{\Lambda _n^{\frac{1}
{{\beta  + 1}}}{\lambda ^{\frac{\beta }
{{\beta  + 1}}}}}} - \frac{1}
{\lambda },{\kern 1pt} {\kern 1pt} {\kern 1pt} {\kern 1pt} {\kern 1pt} {\kern 1pt} {\kern 1pt} {\kern 1pt} {\kern 1pt} {\kern 1pt} \lambda  \geqslant {\Lambda _n}}  \\
   {{\kern 1pt} {\kern 1pt} {\kern 1pt} {\kern 1pt} {\kern 1pt} {\kern 1pt} {\kern 1pt} {\kern 1pt} {\kern 1pt} {\kern 1pt} {\kern 1pt} {\kern 1pt} {\kern 1pt} {\kern 1pt} {\kern 1pt} {\kern 1pt} {\kern 1pt} {\kern 1pt} {\kern 1pt} {\kern 1pt} {\kern 1pt} {\kern 1pt} {\kern 1pt} {\kern 1pt} {\kern 1pt} {\kern 1pt} {\kern 1pt} {\kern 1pt} {\kern 1pt} {\kern 1pt} {\kern 1pt} 0,{\kern 1pt} {\kern 1pt} {\kern 1pt} {\kern 1pt} {\kern 1pt} {\kern 1pt} {\kern 1pt} {\kern 1pt} {\kern 1pt} {\kern 1pt} {\kern 1pt} {\kern 1pt} {\kern 1pt} {\kern 1pt} {\kern 1pt} {\kern 1pt} {\kern 1pt} {\kern 1pt} {\kern 1pt} {\kern 1pt} {\kern 1pt} {\kern 1pt} {\kern 1pt} {\kern 1pt} {\kern 1pt} {\kern 1pt} {\kern 1pt} {\kern 1pt} {\kern 1pt} {\kern 1pt} {\kern 1pt} {\kern 1pt} {\kern 1pt} {\kern 1pt} {\kern 1pt} {\kern 1pt} {\kern 1pt} {\kern 1pt} \lambda  < {\Lambda _n}}  \\

 \end{array} } \right.,
\tag{18}
\end{equation}
where ${\mu _{{\text{opt\_n}}}}(\theta ,\lambda )$ is the optimized transmission power allocated for subchannels in the $n$th group. Therefore, the optimized energy efficiency  of MIMO-OFDM mobile multimedia communication systems with statistical QoS constraints is derived as (19) and (20).

\subsection{Algorithm Design}
\label{sec4-2}
The core idea of energy efficiency optimization algorithm (EEOPA) with statistical QoS constraints for MIMO-OFDM mobile multimedia communication systems is described as follows. Firstly, the SVD method is applied for the channel matrix ${{\mathbf{H}}_k}$, $k = 1,2,...,N$, at each orthogonal subcarrier to obtain $M \times N$ parallel space-frequency subchannels. Secondly, subchannels at each subcarrier are pushed into a subchannel gain set, where subchannels are sorted by the subchannel gain in a descending order. And then subchannles with the same order position in the subchannel gain set are selected into the same group. Since subchannels within the same group have the identical MPDF, the transmission power allocation threshold for subchannels within the same group is identical. Therefore, the optimized transmission power allocation for the grouped subchannels is implemented to improve the energy efficiency of MIMO-OFDM mobile multimedia communication systems.
The detailed EEOPA algorithm is illustrated in Algorithm~\ref{alg1}.

\begin{algorithm*}
\centering
\renewcommand{\algorithmicrequire}{\textbf{Input:}}
\renewcommand\algorithmicensure {\textbf{Output:} }
\begin{algorithmic}
\caption{EEOPA.}
\label{alg1}

\REQUIRE ${M_t}$, ${M_r}$, $N$, ${{\mathbf{H}}_k}$, $\overline{P}$,$B$,${T_f}$,$\theta $;\\
\STATE \textbf{Initialization:} Decompose the MIMO-OFDM channel matrix ${{\mathbf{H}}_k} (k = 1,2,...,N)$ into $M \times N $ space-frequency subchannels through the SVD method.
\STATE \textbf{Begin:} \begin{enumerate}
                         \item Sort subchannel gains of each subcarrier in a decreasing order:
                         \begin{equation}\label{eq21}
                           {\lambda _{1,k}} \geqslant {\lambda _{2,k}} \geqslant ... \geqslant {\lambda _{M,k}}(k = 1,2,...,N).
                         \tag{21}
                         \end{equation}
                         \item Assign ${\lambda _{n,1}},{\lambda _{n,2}},...,{\lambda _{n,N}}$ from all $N$ subcarriers into the $n$th grouped subchannel set:
                         \begin{equation}\label{eq22}
                           {\text{Group\_}}\operatorname{n}  = \{ {\lambda _{n,1}},{\lambda _{n,2}},...,{\lambda _{n,N}}\} (n = 1,2,...,M).
                         \tag{22}
                         \end{equation}
                         \item \textbf{for} {$n = 1:M$} \textbf{do} \\
                          \quad Calculate the optimized transmission power allocation threshold ${\Lambda _n}$ for ${\text{Group\_n}}$ \\
                          \quad according to the average power constraint as follows:
                           \begin{equation}\label{eq23}
                             \int_{{\Lambda _n}}^\infty  {(\frac{1}{{\Lambda _n^{\frac{1}{{\beta  + 1}}}{\lambda ^{\frac{\beta }{{\beta  + 1}}}}}} - \frac{1}{\lambda }){p_{\Gamma n}}(\lambda )d\lambda  \le \overline {P}}.
                           \tag{23}
                           \end{equation}
                           \quad Execute the optimized transmission power allocation policy for ${\text{Group\_n}}$:
                           \begin{equation}\label{eq24}
                             {\mu _{{\text{opt\_n}}}}(\theta ,\lambda ) = \left\{ {\begin{array}{*{20}{c}}{\frac{1}{{\Lambda _n^{\frac{1}{{\beta  + 1}}}{\lambda^{\frac{\beta }{{\beta  + 1}}}}}} - \frac{1}{\lambda },{\kern 1pt} {\kern 1pt} {\kern 1pt} {\kern 1pt} {\kern 1pt} {\kern 1pt} {\kern 1pt} {\kern 1pt} {\kern 1pt} {\kern 1pt} \lambda  \geqslant {\Lambda _n}}  \\{{\kern 1pt} {\kern 1pt} {\kern 1pt} {\kern 1pt} {\kern 1pt} {\kern 1pt} {\kern 1pt} {\kern 1pt} {\kern 1pt} {\kern 1pt} {\kern 1pt} {\kern 1pt} {\kern 1pt} {\kern 1pt} {\kern 1pt} {\kern 1pt} {\kern 1pt} {\kern 1pt} {\kern 1pt} {\kern 1pt} {\kern 1pt} {\kern 1pt} {\kern 1pt} {\kern 1pt} {\kern 1pt} {\kern 1pt} {\kern 1pt} {\kern 1pt} {\kern 1pt} {\kern 1pt} {\kern 1pt} 0,{\kern 1pt} {\kern 1pt} {\kern 1pt} {\kern 1pt} {\kern 1pt} {\kern 1pt} {\kern 1pt} {\kern 1pt} {\kern 1pt} {\kern 1pt} {\kern 1pt} {\kern 1pt} {\kern 1pt} {\kern 1pt} {\kern 1pt} {\kern 1pt} {\kern 1pt} {\kern 1pt} {\kern 1pt} {\kern 1pt} {\kern 1pt} {\kern 1pt} {\kern 1pt} {\kern 1pt} {\kern 1pt} {\kern 1pt} {\kern 1pt} {\kern 1pt} {\kern 1pt} {\kern 1pt} {\kern 1pt} {\kern 1pt} {\kern 1pt} {\kern 1pt} {\kern 1pt} {\kern 1pt} {\kern 1pt} {\kern 1pt} \lambda  < {\Lambda _n}}  \\\end{array} } \right..
                           \tag{24}
                           \end{equation}
                           \quad Calculate the optimized effective-capacity for ${\text{Group\_n}}$:
                           \begin{equation}\label{eq25}
                             {C_e}{(\theta )_{{\text{opt\_}}\operatorname{n} }} =  - \frac{N}{\theta }\log \left( {\int_0^\infty  {{e^{ - \theta {T_f}B{{\log }_2}(1 + {\mu _{{\text{opt\_}}\operatorname{n} }}(\theta ,\lambda )\lambda )}}{p_{\Gamma n}}(\lambda )d\lambda } } \right).
                           \tag{25}
                           \end{equation}
                           \textbf{end for}
                         \item Calculate the optimized energy-efficiency of the MIMO-OFDM mobile multimedia communication system:
                         \begin{equation}\label{eq26}
                           {\eta _{\text{opt}}} =  - \frac{1}{{\theta  \times \mathop {P \times }\limits^ -  M}}\sum\limits_{n = 1}^M {\log \left( {\int_0^\infty  {{e^{ - \theta {T_f}B{{\log }_2}(1 + {\mu _{\operatorname{opt} \_n}}(\theta ,\lambda )\lambda )}}{p_{{\Gamma _n}}}(\lambda )d\lambda } } \right)}.
                         \tag{26}
                         \end{equation}
                       \end{enumerate}
\STATE \textbf{end Begin}
\ENSURE ${\Lambda _n}$, ${\eta _{{\text{opt}}}}$.\\ 
\end{algorithmic}
\end{algorithm*}

\section{Simulation Results and Performance Analysis}
\label{sec5}
In the proposed algorithm, the transmission power allocation threshold ${\Lambda _n}$ is the core parameter to optimize the energy efficiency of MIMO-OFDM mobile multimedia communication systems. The configuration of the transmission power allocation threshold ${\Lambda _n}$ depends on the MPDF of each grouped subchannels. Without loss of generation, the number of transmitter and receiver antennas is configured as ${M_t} = 4$ and ${M_r} = 4$, respectively. Based on the extension of (\ref{eq16}), MPDFs of each grouped subchannels are extended as (27)-(30).
\begin{figure*}[!t]
\begin{equation}\label{eq27}
\begin{split}
  {p_{{\Gamma _1}}}(\lambda ) =  &- 4{e^{ - 4\lambda }} - (1/36){e^{ - \lambda }}(144 - 432\lambda  + 648{\lambda ^2} - 408\lambda _{}^3 +  126\lambda _{}^4 \hfill \\
  &- 18\lambda _{}^5 + \lambda _{}^6) + (1/12){e^{ - 3\lambda }}(144 - 144{\lambda _{}} + 72\lambda _{}^2 + 56\lambda _{}^3 + 46\lambda _{}^4 \hfill \\
  &+ 10\lambda _{}^5 + \lambda _{}^6) - (1/72){e^{ - 2{\lambda _{}}}}(864 - 1728{\lambda _{}} + 1728\lambda _{}^2 - 192\lambda _{}^3 \hfill \\
   &+ 96\lambda _{}^4 - 96\lambda _{}^5 + 32\lambda _{}^6 - 4\lambda _{}^7 + \lambda _{}^8),\hfill \\
\end{split}
\tag{27}
\end{equation}
\end{figure*}
\begin{figure*}[!t]
\begin{equation}\label{eq28}
\begin{split}
  {p_{{\Gamma _2}}}(\lambda ) = &12{e^{ - 4\lambda }} - (1/6){e^{ - 3\lambda }}(144 - 144\lambda  + 72\lambda _{}^2 + 56\lambda _{}^3 + 46\lambda _{}^4 \hfill \\
   &+ 10\lambda _{}^5 + \lambda _{}^6) + (1/72){e^{ - 2\lambda }}(864 - 1728\lambda  + 1728\lambda _{}^2 - 192\lambda _{}^3  \hfill \\
   &+ 96\lambda _{}^4- 96\lambda _{}^5 + 32\lambda _{}^6 - 4\lambda _{}^7 + \lambda _{}^8), \hfill \\
\end{split}
\tag{28}
\end{equation}
\end{figure*}
\begin{figure*}[!t]
\begin{equation}\label{eq29}
  \begin{split}
  {p_{{\Gamma _3}}}(\lambda ) =  &- 12{e^{ - 4\lambda }} + (1/12){e^{ - 3\lambda }}(144 - 144\lambda  + 72\lambda _{}^2 + 56\lambda _{}^3 + 46\lambda _{}^4 \hfill \\
   &+ 10\lambda _{}^5 + \lambda _{}^6), \hfill \\
\end{split}
\tag{29}
\end{equation}
\end{figure*}
\begin{figure*}[!t]
\begin{equation}\label{eq30}
  {p_{{\Gamma _4}}}(\lambda ) = 4{e^{ - 4\lambda }}.
\tag{30}
\end{equation}
\end{figure*}

\begin{figure}
\vspace{0.1in}
\centerline{\includegraphics[width=9cm,draft=false]{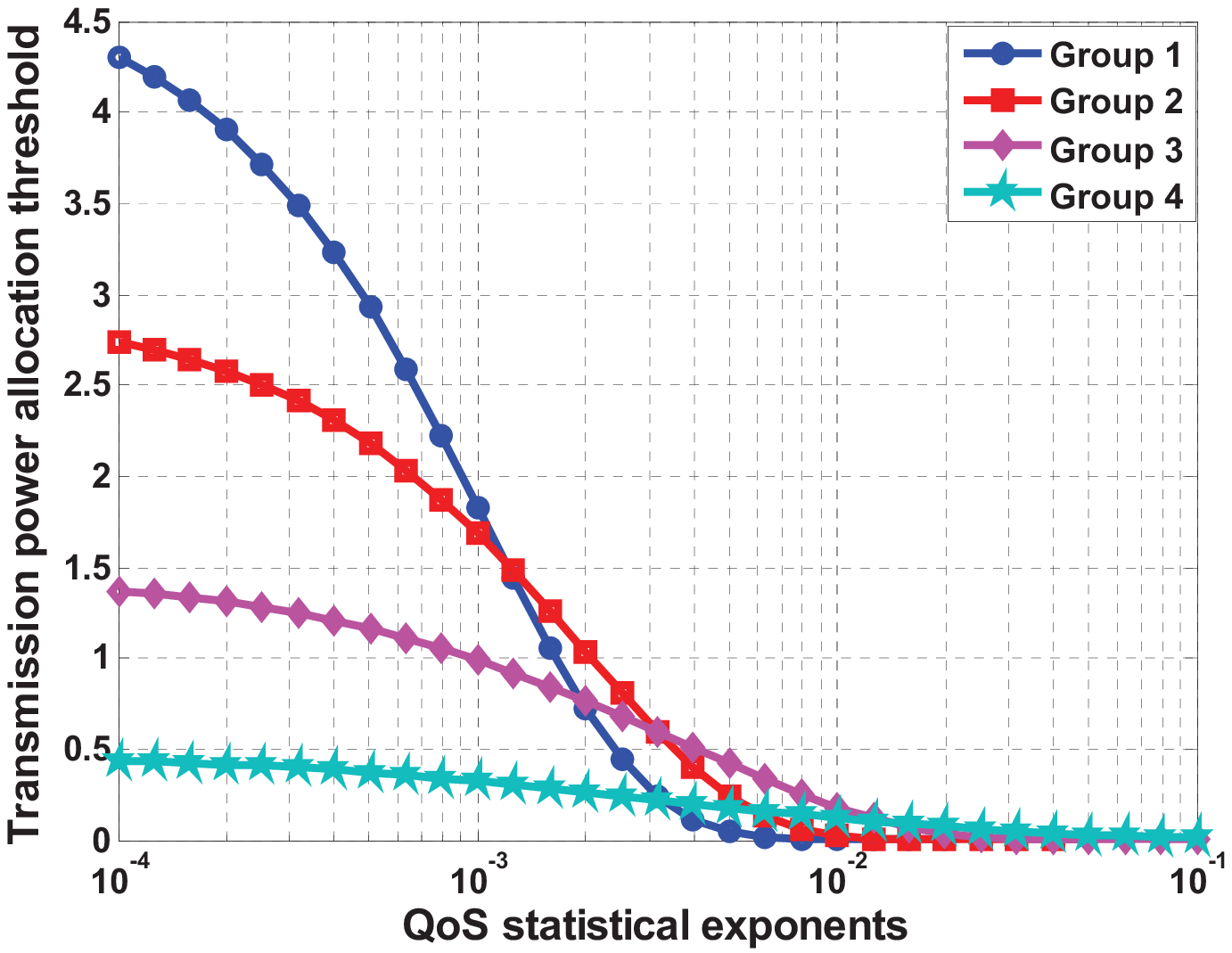}}
\caption{\small Transmission power allocation threshold ${\Lambda _n}$ with respect to each grouped subchannels considering different QoS statistical exponents $\theta$.}
\label{fig2}
\end{figure}

\begin{figure}
\vspace{0.1in}
\centerline{\includegraphics[width=9cm,draft=false]{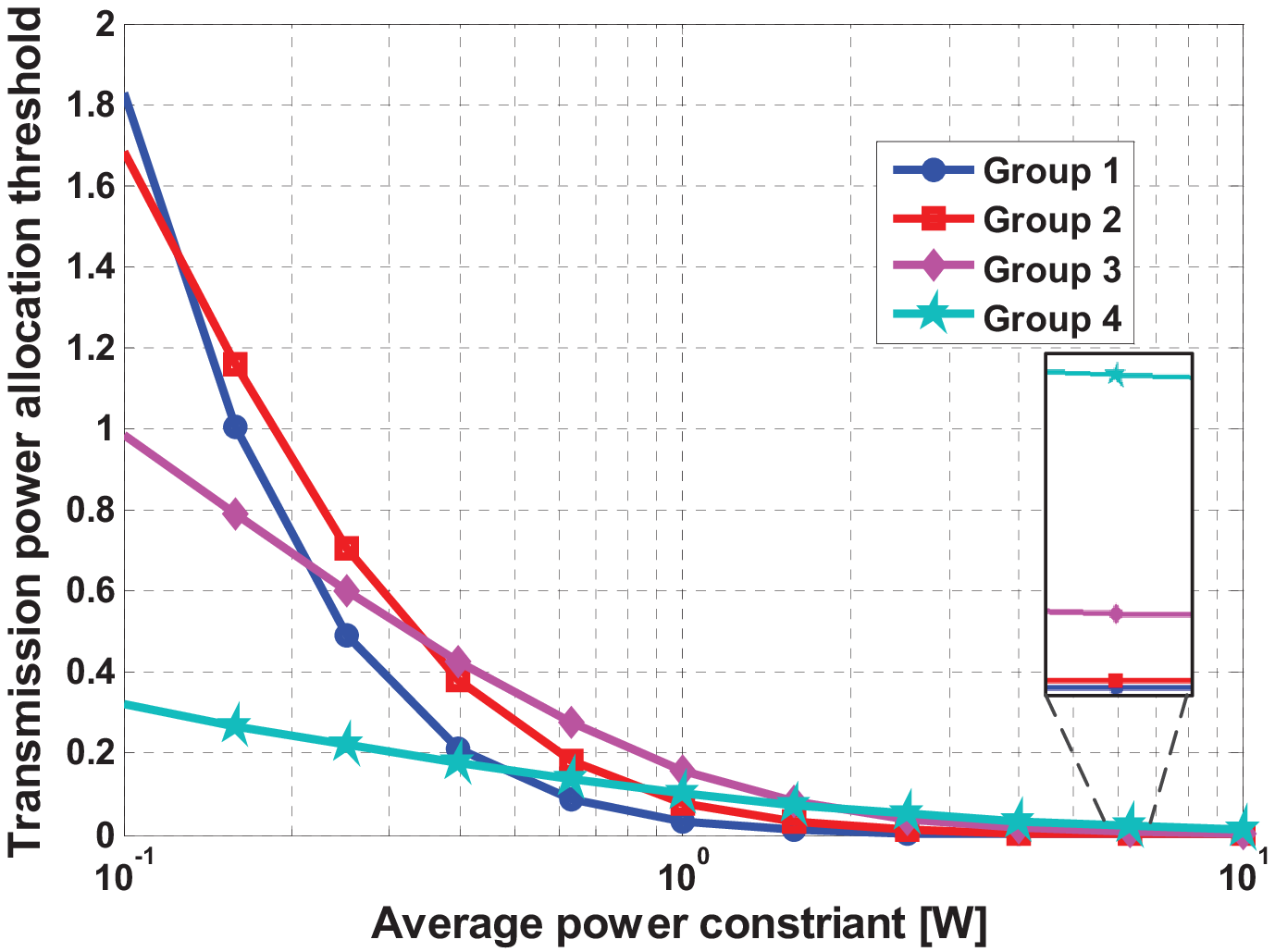}}
\caption{\small Transmission power allocation threshold ${\Lambda _n}$ with respect to each grouped subchannels considering different average power constraints ${\overline {P}}$.}
\label{fig3}
\end{figure}

Substituting (\ref{eq27}), (\ref{eq28}), (\ref{eq29}) and (\ref{eq30}) into (\ref{eq12}), the transmission power allocation threshold ${\Lambda _n}$ can be calculated. To analyze the performance of the transmission power allocation threshold, some default parameters are configured as: ${T_f} = 1ms$ and $B = 1MHz$. The numerical results are illustrated in Fig.~\ref{fig2} and Fig.~\ref{fig3}.
Fig.~\ref{fig2} shows numerical results of the transmission power allocation threshold ${\Lambda _n}$ with respect to each grouped subchannels considering different QoS statistical exponents $\theta$. For each grouped subchannels, the transmission power allocation threshold ${\Lambda _n}$ decreases with the increase of the QoS exponent $\theta$. Considering subchannels are sorted by the descending order of subchannel gains, the subchannel gain of subchannel groups decreases with the increase of group indexes. Therefore, the transmission power allocation threshold ${\Lambda _n}$ increases with the increase of subchannel gains in subchannel groups when the QoS exponent $\theta  \leqslant {10^{ - 3}}$. When the QoS exponent $\theta  > {10^{ - 3}}$, the transmission power allocation threshold ${\Lambda _n}$ start to decrease with the increase of subchannel gains in subchannel groups.

Fig.~\ref{fig3} illustrates the transmission power allocation threshold ${\Lambda _n}$ with respect to each grouped subchannels considering different average power constraints $\overline{P}$. For each grouped subchannels, the transmission power allocation threshold ${\Lambda _n}$ decreases with the increase of the average power constraint $\overline{P}$. When $\overline{P} \leqslant {0.13}$, the transmission power allocation threshold ${\Lambda _n}$ increases with the increase of subchannel gains in subchannel groups. When $\overline{P} > {0.13}$, the transmission power allocation threshold ${\Lambda _n}$ start to decrease with the increase of subchannel gains in subchannel groups.

\begin{figure}
\vspace{0.1in}
\centerline{\includegraphics[width=9cm,draft=false]{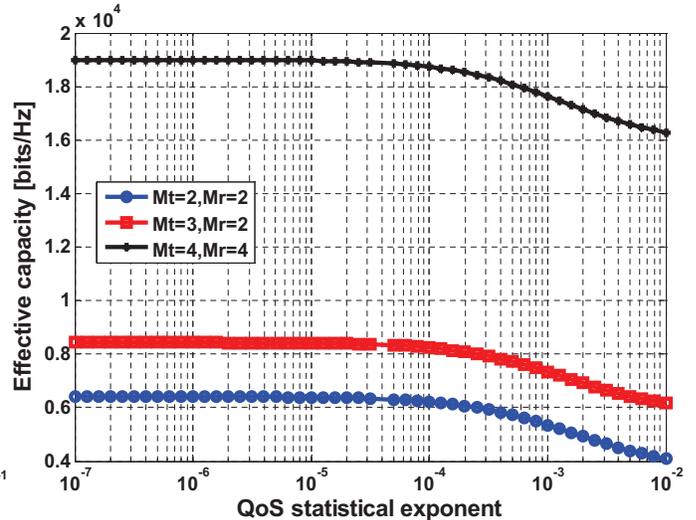}}
\caption{\small  Effective capacity ${C_{total}}(\theta )$ with respect to the QoS statistical exponent $\theta$ considering different scenarios.}
\label{fig4}
\end{figure}
\begin{figure}
\vspace{0.1in}
\centerline{\includegraphics[width=9cm,draft=false]{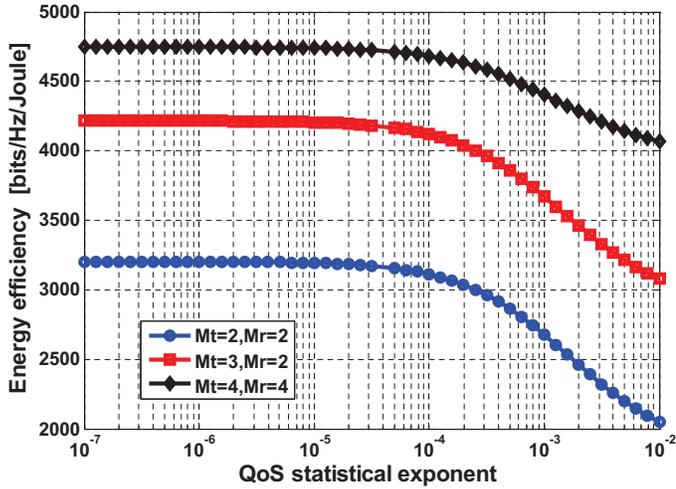}}
\caption{\small  Energy efficiency $\eta $ with respect to the QoS statistical exponent $\theta$ considering different scenarios.}
\label{fig5}
\end{figure}

To evaluate the energy efficiency and the effective capacity of MIMO-OFDM mobile multimedia communication systems, three typical scenarios with different antenna numbers are configured in Fig.~\ref{fig4} and Fig.~\ref{fig5}: (1) ${M_t = 2}$, ${M_r = 2}$; (2) ${M_t = 3}$, ${M_r = 2}$; (3) ${M_t = 4}$, ${M_r = 4}$. Fig.~\ref{fig4} shows the impact of QoS statistical exponents $\theta$ on the effective capacity of MIMO-OFDM mobile multimedia communication systems in three different scenarios. From curves in Fig.~\ref{fig4}, the effective capacity decreases with the increase of the QoS statistical exponent $\theta$. The reason of this result is that the larger values of $\theta$ correspond to the higher QoS requirements, which result in a smaller number of subchannels are selected to satisfy the higher QoS requirements. When the QoS statistical exponent $\theta$ is fixed, the effective capacity increases with the number of antennas in MIMO-OFDM mobile multimedia communication systems. This result indicates the channel spatial multiplexing can improve the effective capacity of MIMO-OFDM mobile multimedia communication systems.

Fig.~\ref{fig5} illustrates the impact of QoS statistical exponents $\theta$ on the energy efficiency of MIMO-OFDM mobile multimedia communication systems in three different scenarios. From curves in Fig.~\ref{fig5}, the energy efficiency decreases with the increase of the QoS statistical exponent $\theta$. The reason of this result is that the larger values of $\theta$ correspond to the higher QoS requirements, which result in a smaller number of subchannels are selected to satisfy the higher QoS requirements. This result conduces to the effective capacity is decreased. If the total transmission power is constant, the decreased effective capacity will lead to the decrease of the energy efficiency in communication systems. When the QoS statistical exponent $\theta$ is fixed, the energy efficiency increases with the number of antennas in MIMO-OFDM mobile multimedia communication systems. This result indicates the channel spatial multiplexing can improve the energy efficiency of MIMO-OFDM mobile multimedia communication systems.

\begin{figure}
\vspace{0.1in}
\centerline{\includegraphics[width=9cm,draft=false]{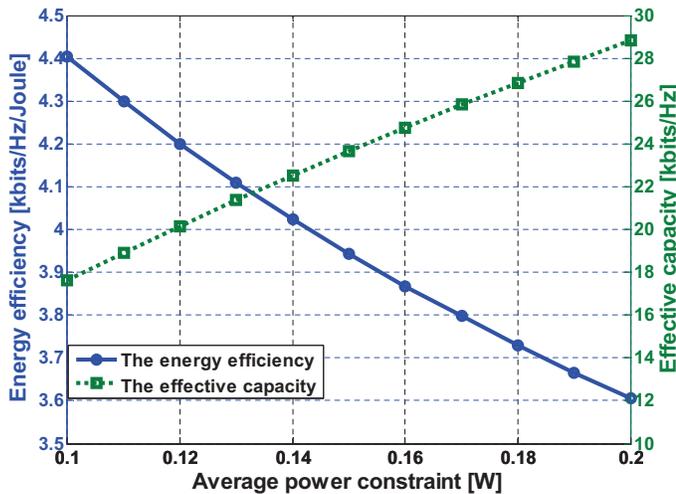}}
\caption{\small  Impact of the average power constraint on the energy efficiency $\eta $ and the effective capacity ${C_{total}}(\theta )$.}
\label{fig6}
\end{figure}

When the QoS statistical exponent is fixed as $\theta  = {10^{ - 3}}$, the impact of the average power constraint on the energy efficiency and the effective capacity of MIMO-OFDM mobile multimedia communication systems is investigated in Fig.~\ref{fig6}. From Fig.~\ref{fig6}, the energy efficiency decreases with the increase of the average power constraint and the affective capacity increases with the increase of the average power constraint. This result implies there is an optimization tradeoff between the energy efficiency and effective capacity in MIMO-OFDM mobile multimedia communication systems:as the transmission power increases which leads to larger effective capacity, the energy consumption of the system also rises; therefor, the larger power input results in the decline of energy efficiency.

\begin{figure}
\vspace{0.1in}
\centerline{\includegraphics[width=9cm,draft=false]{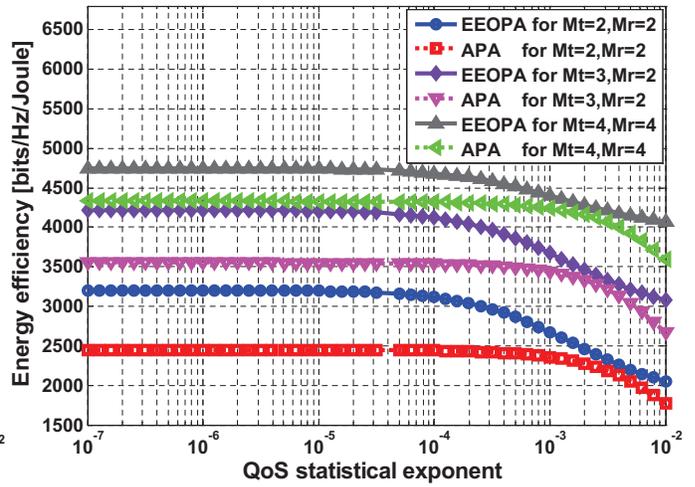}}
\caption{\small  Energy efficiency $\eta $ of EEOPA and APA algorithms as variation of QoS statistical exponent $\theta$ under different scenarios.}
\label{fig7}
\end{figure}

\begin{figure}
\vspace{0.1in}
\centerline{\includegraphics[width=9cm,draft=false]{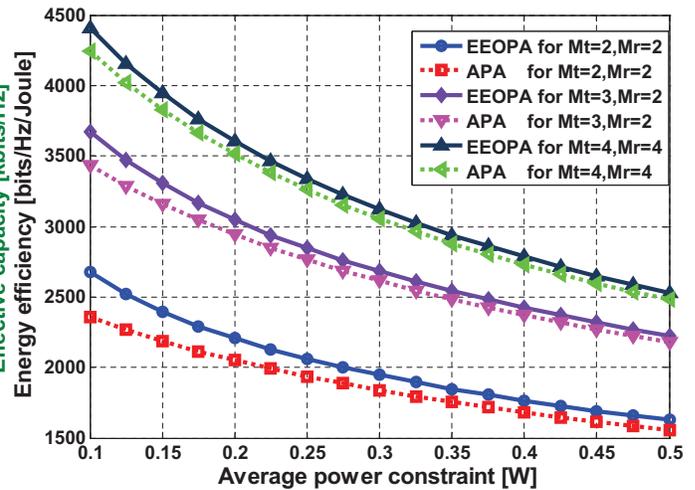}}
\caption{\small  Energy efficiency $\eta $ of EEOPA and APA algorithms as variation of average power constraint $\overline {P}$ under different scenarios.}
\label{fig8}
\end{figure}
\begin{figure}
\vspace{0.1in}
\centerline{\includegraphics[width=9cm,draft=false]{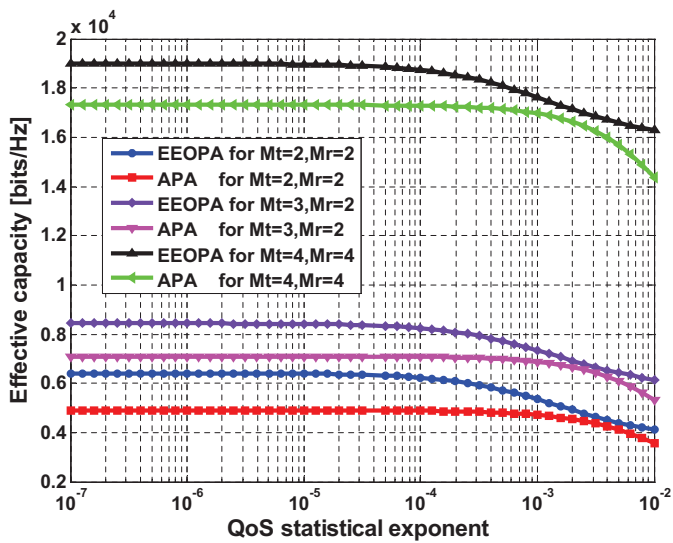}}
\caption{\small   Effective capacity ${C_{total}}(\theta )$ of EEOPA and APA algorithms as variation of QoS statistical exponent $\theta$ under different scenarios.}
\label{fig9}
\end{figure}
\begin{figure}
\vspace{0.1in}
\centerline{\includegraphics[width=9cm,draft=false]{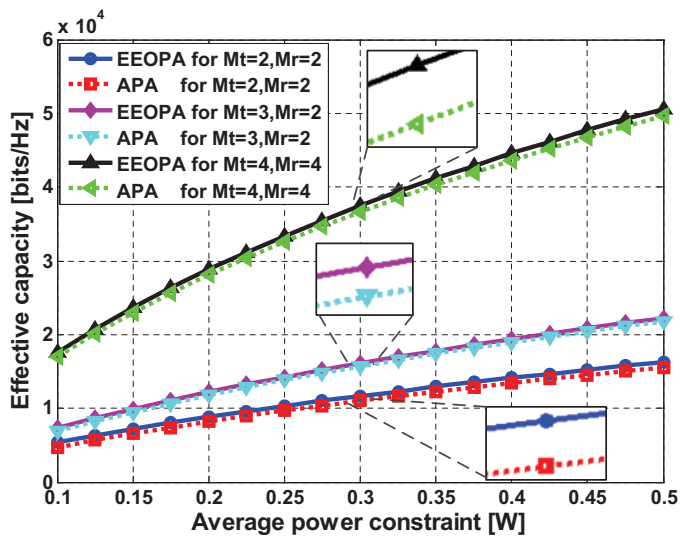}}
\caption{\small Effective capacity ${C_{total}}(\theta )$ of EEOPA and APA algorithm as variation of average power constraint $\overline {P}$ under different scenarios.}
\label{fig10}
\end{figure}

To analyze performance of the EEOPA algorithm, the traditional average power allocation (APA) algorithm \cite{Zhihua08}, i.e., every subchannel with the equal transmission power algorithm is compared with the EEOPA algorithm by Fig.~\ref{fig7}--Fig.~\ref{fig10}. Three typical scenarios with different antenna numbers are configured in Fig.~\ref{fig7}--Fig.~\ref{fig10}: (1) $M_t = 2$, $M_r = 2$; (2) $M_t = 3$, $M_r = 2$; (3) $M_t = 4$, $M_r = 4$. In Fig.~\ref{fig7}, the effect of the QoS statistical exponent $\theta$ on the energy efficiency of EEOPA and APA algorithms is investigated with constant average power constraint ${\overline{P} = 0.1}$ Watt. Considering changes of the QoS statistical exponent, the energy efficiency of EEOPA algorithm is always higher than the energy efficiency of APA algorithm in three scenarios. In Fig.~\ref{fig8}, the impact of the average power constraint on the energy efficiency of EEOPA and APA algorithms is evaluated with the fixed QoS statistical exponent $\theta  = {10^{ - 3}}$. Considering changes of the average power constraint, the energy efficiency of EEOPA algorithm is always higher than the energy efficiency of APA algorithm in three scenarios.
In Fig.~\ref{fig9}, the effect of the QoS statistical exponent $\theta$ on the effective capacity of EEOPA and APA algorithms is compared with constant average power constraint ${\overline{P} = 0.1}$ Watt. Considering changes of the QoS statistical exponent, the effective capacity of EEOPA algorithm is always higher than the effective capacity of APA algorithm in three scenarios. In Fig.~\ref{fig10}, the impact of the average power constraint on the effective capacity of EEOPA and APA algorithms is evaluated with the fixed QoS statistical exponent $\theta  = {10^{ - 3}}$. Considering changes of the average power constraint, the effective capacity of EEOPA algorithm is always higher than the effective capacity of APA algorithm in three scenarios. Based on above comparison results, our proposed EEOPA algorithm can improve the energy efficiency and effective capacity of MIMO-OFDM mobile multimedia communication systems.

\section{Conclusions}
\label{sec6}
In this paper, an energy efficiency model is proposed for MIMO-OFDM mobile multimedia communication systems with statistical QoS constraints. An energy efficiency optimization scheme is presented based on the subchannel grouping method, in which the complex multi-channel joint optimization problem is simplified into a multi-target single channel optimization problem. A closed-form solution of the energy efficiency optimization is derived for MIMO-OFDM mobile multimedia communication systems. Moreover, a novel algorithm, i.e., EEOPA, is designed to improve the energy efficiency of MIMO-OFDM mobile multimedia communication systems. Compared with the traditional APA algorithm, simulation results demonstrate that our proposed algorithm has advantages on improving the energy efficiency and effective capacity of MIMO-OFDM mobile multimedia communication systems with QoS constraints.


\begin{biography}[{\includegraphics[width=1in,height=1.25in,clip,keepaspectratio]{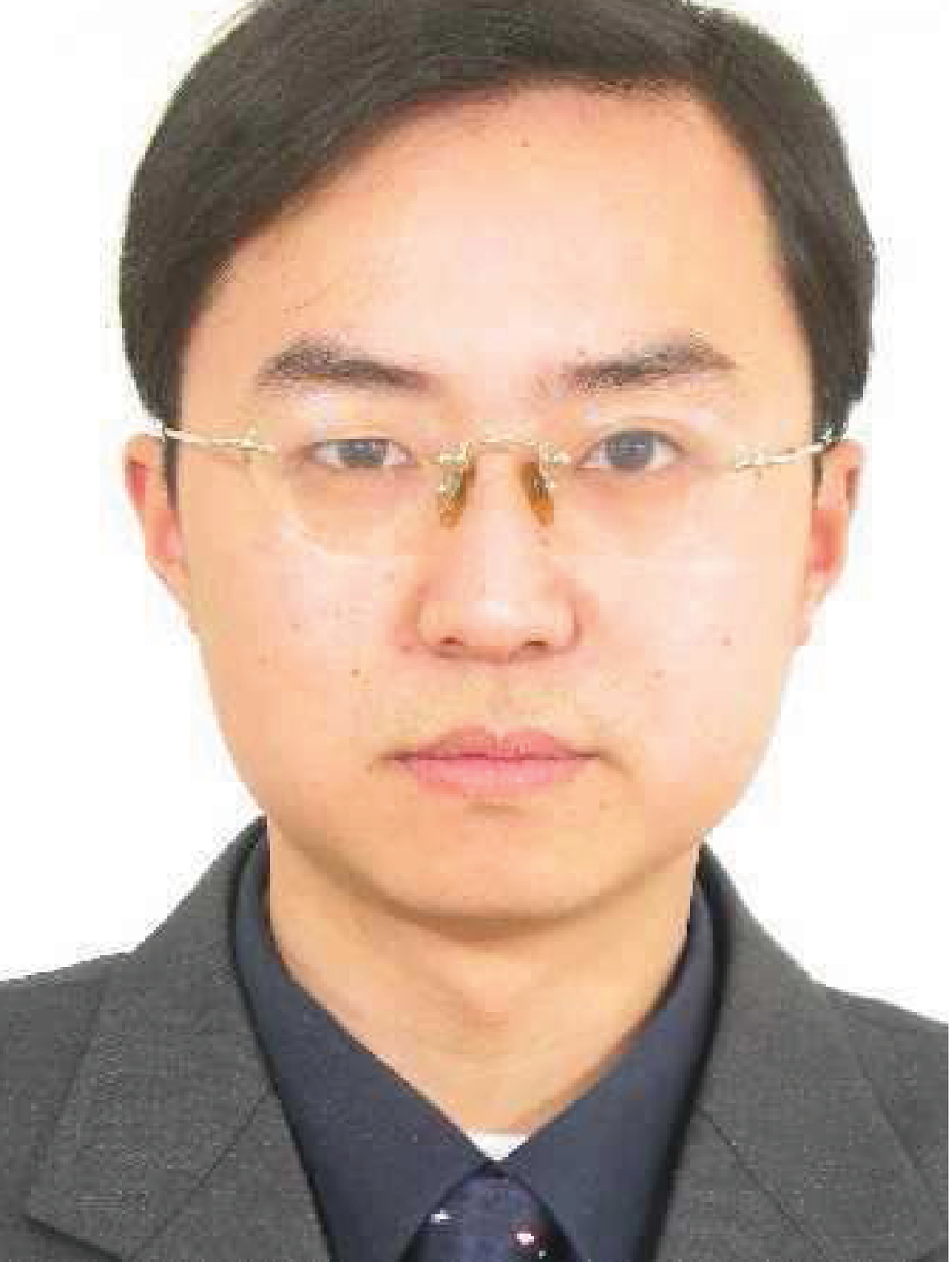}}]{Xiaohu~Ge}
(M'09-SM'11) is currently a Professor with the Department of Electronics and Information Engineering at Huazhong University of Science and Technology (HUST), China. He received his PhD degree in Communication and Information Engineering from HUST in 2003. He has worked at HUST since Nov. 2005. Prior to that, he worked as a researcher at Ajou University (Korea) and Politecnico Di Torino (Italy) from Jan. 2004 to Oct. 2005. He was a visiting researcher at Heriot-Watt University, Edinburgh, UK from June to August 2010. His research interests are in the area of mobile communications, traffic modeling in wireless networks, green communications, and interference modeling in wireless communications. He has published about 60 papers in refereed journals and conference proceedings and has been granted about 15 patents in China. He received the Best Paper Awards from IEEE Globecom 2010. He is leading several projects funded by NSFC, China MOST, and industries. He is taking part in several international joint projects, such as the RCUK funded UK-China Science Bridges: R\&D on (B)4G Wireless Mobile Communications and the EU FP7 funded project: Security, Services, Networking and Performance of Next Generation IP-based Multimedia Wireless Networks.

Dr. Ge is currently serving as an Associate Editor for \textit{International Journal of Communication Systems (John Wiley \& Sons)}, \textit{IET Wireless Sensor Systems}, \textit{KSII Transactions on Internet and Information Systems} and \textit{Journal of Internet Technology}. Since 2005, he has been actively involved in the organisation of more than 10 international conferences, such as Executive Chair of IEEE GreenCom 2013 and Co-Chair of workshop of Green Communication of Cellular Networks at IEEE GreenCom 2010. He is a Senior Member of the IEEE, a Senior member of the Chinese Institute of Electronics, a Senior member of the China Institute of Communications, and a member of the NSFC and China MOST Peer Review College.\end{biography}

\begin{biography}[{\includegraphics[width=1in,height=1.25in,clip,keepaspectratio]{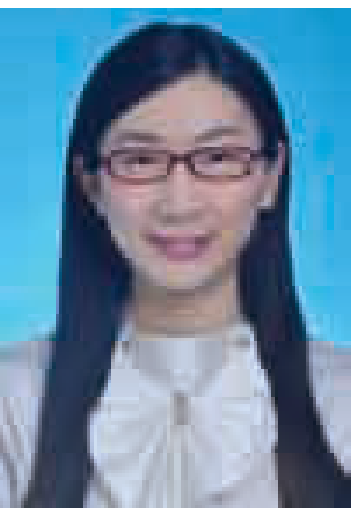}}]{Xi~Huang}
received her Bachelor's degree in telecommunication engineering from Huazhong University of Science and Technology (HUST) in June 2011. She is currently working toward her Master's degree in communication and information systems at HUST. Her research interests include energy efficiency modeling and performance analysis in wireless communication systems.\end{biography}

\begin{biography}[{\includegraphics[width=1in,height=1.25in,clip,keepaspectratio]{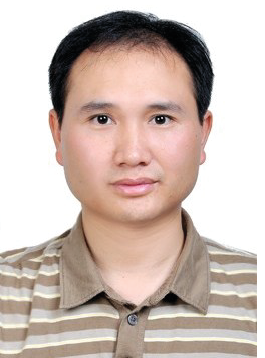}}]{Yuming~Wang}
received his Ph.D. degree in Communication and Information Engineering from Huazhong University of Science and Technology, China in 2005. He is currently an associate professor with the Department of Electronics and Information Engineering, Huazhong University of Science and Technology, China. His research interests include wired/wireless communications, vehicular networks, mobile applications, database and data processing. Since 1999, he had been working in the design of high speed network router and switch products for nearly 10 years. In the recent years, he has been researching on energy efficiency modeling in wireless communications, mobile IP networks, vehicular ad hoc networks, software application system, statistical analysis and data mining. He has published over 10 papers in refered journals and conference proceedings and has been granted several patents in China. He is leading projects funded by NSFC and MOE of China.\end{biography}

\begin{biography}[{\includegraphics[width=1in,height=1.25in,clip,keepaspectratio]{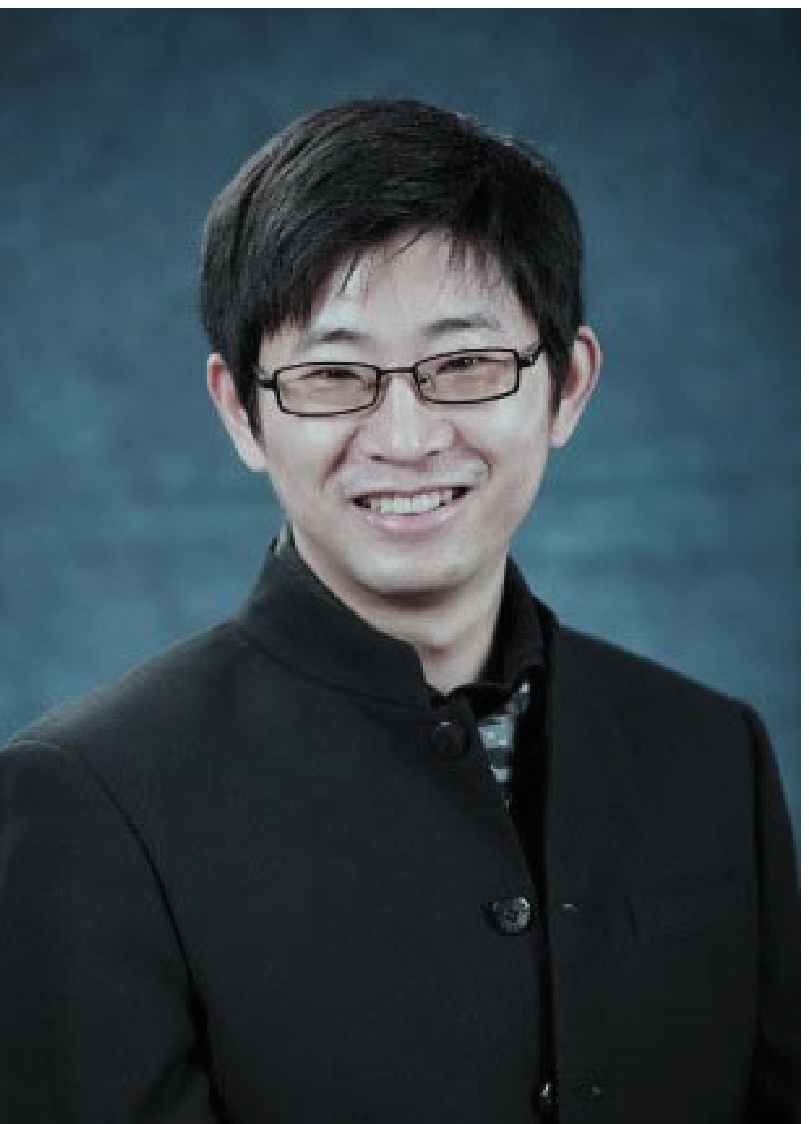}}]{Min~Chen}
 is a professor in School of Computer Science and Technology at Huazhong University of Science and Technology (HUST). He was an assistant professor in School of Computer Science and Engineering at Seoul National University (SNU) from Sep. 2009 to Feb. 2012. He was R\&D director at Confederal Network Inc. from 2008 to 2009. He worked as a Post-Doctoral Fellow in Department of Electrical and Computer Engineering at University of British Columbia (UBC) for three years. Before joining UBC, he was a Post-Doctoral Fellow at SNU for one and half years. He received Best Paper Award from IEEE ICC 2012, and Best Paper Runner-up Award from QShine 2008. He has more than 180 paper publications, including 85 SCI papers. He is a Guest Editor for IEEE Network, IEEE Wireless Communications Magazine, etc. He is Co-Chair of IEEE ICC 2012-Communications Theory Symposium, and Co-Chair of IEEE ICC 2013-Wireless Networks Symposium. He is General Co-Chair for the 12th IEEE International Conference on Computer and Information Technology (IEEE CIT-2012). He is Keynote Speaker for CyberC 2012 and Mobiquitous 2012. His research focuses on Internet of Things, Machine to Machine Communications, Body Area Networks, Body Sensor Networks, E-healthcare, Mobile Cloud Computing, Cloud-Assisted Mobile Computing, Ubiquitous Network and Services, Mobile Agent, and Multimedia Transmission over Wireless Network, etc.\end{biography}

\begin{biography}[{\includegraphics[width=1in,height=1.25in,clip,keepaspectratio]{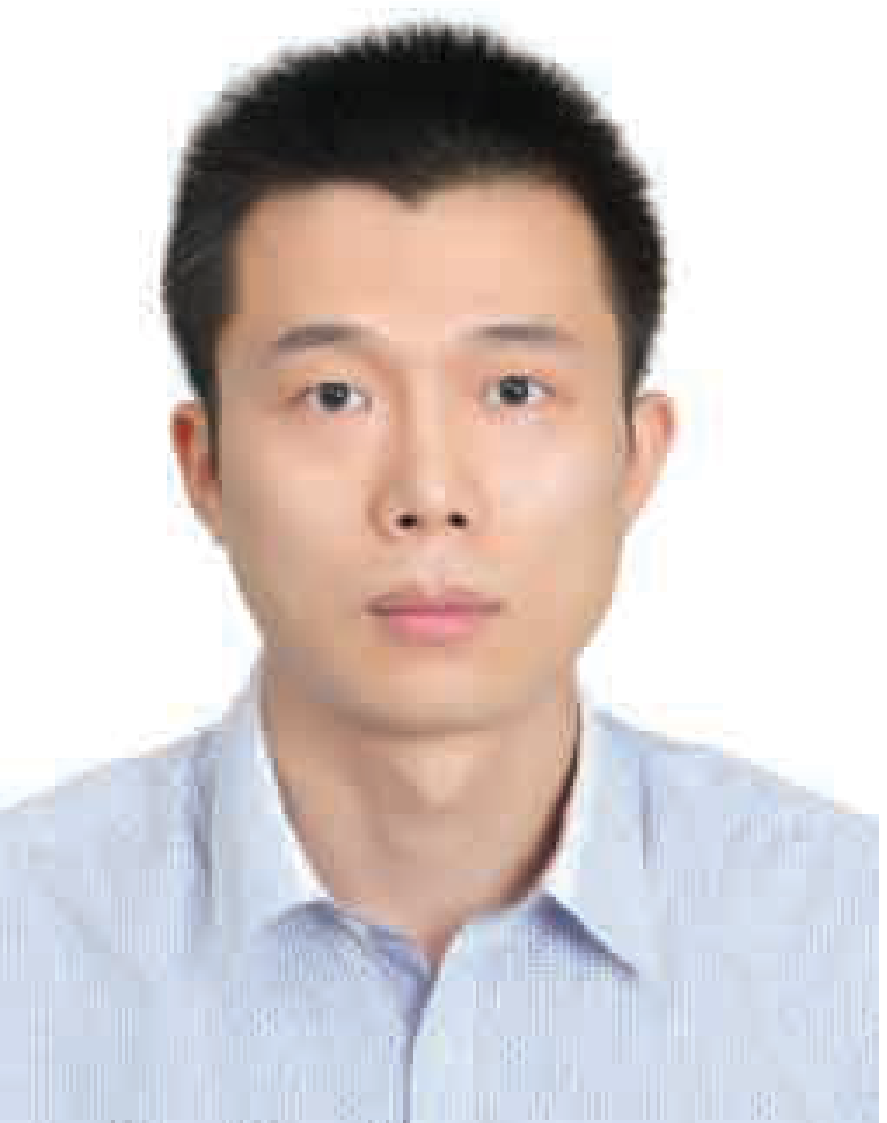}}]{Qiang~Li}
received his B.Eng degree in communication engineering from the School of Communication and Information Engineering, University
of Electronic Science and Technology of China, in 2007, and the Ph.D. degree in electrical and electronic engineering from Nanyang Technological University, in 2011. From 2011-2013, he was with Nanyang Technological University as a Research Fellow. From 2013, he is an Associate Professor at Huazhong University of Science and Technology. His current research interests include cooperative communications, cognitive wireless networks, and wireless network coding .
\end{biography}

\begin{biography}[{\includegraphics[width=1in,height=1.25in,clip,keepaspectratio]{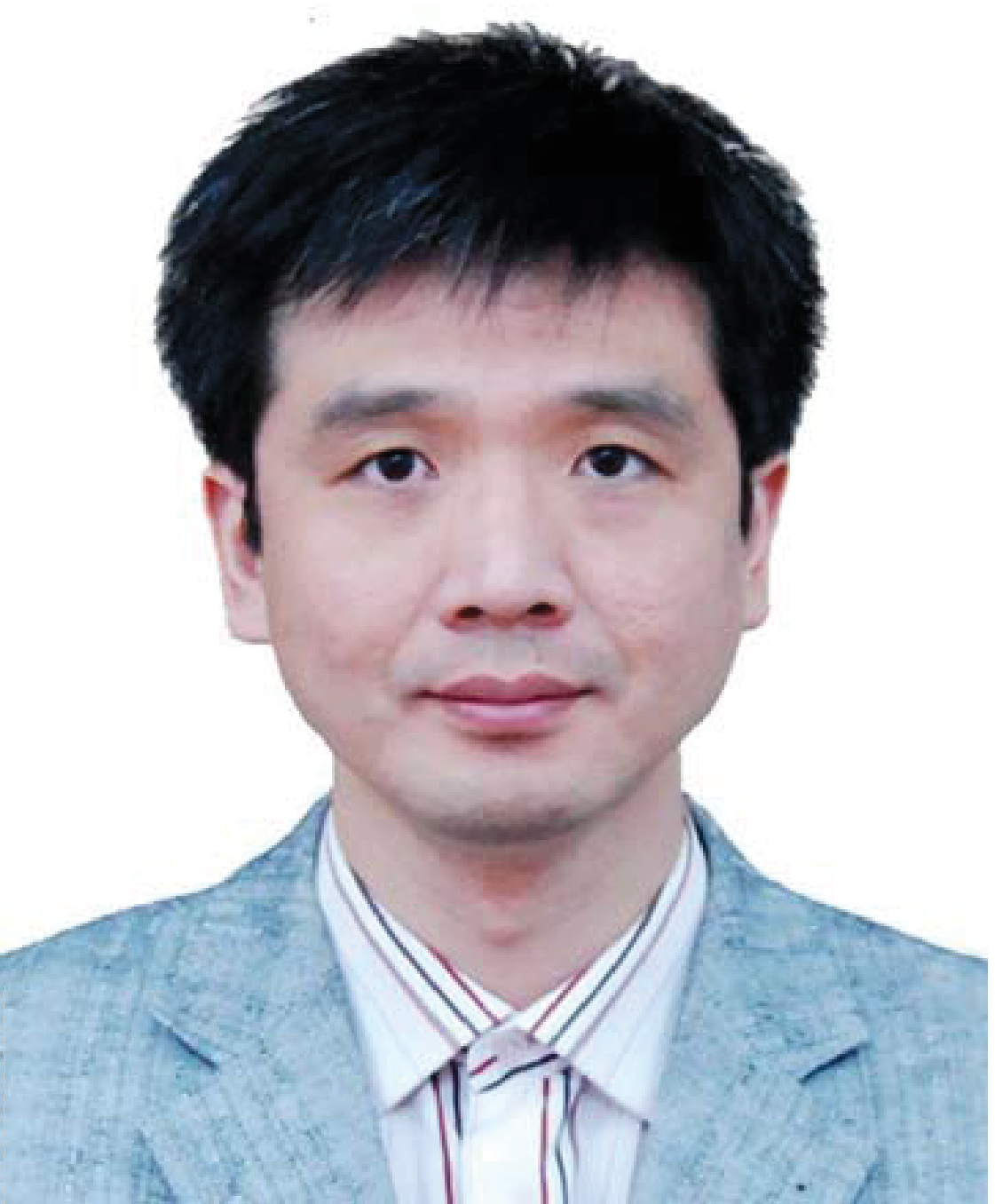}}]{Tao~Han}
received the Ph.D. degree in communication and information engineering from Huazhong University of Science and Technology (HUST), Wuhan, China in December, 2001. He is currently an Associate Professor with the Department of Electronics and Information Engineering, HUST. From August, 2010 to August, 2011, he was a Visiting Scholar with University of Florida, Gainesville, FL, USA, as a Courtesy Associate Professor. His research interests include wireless communications, multimedia communications, and computer networks.
Dr. Han is currently serving as an Area Editor for the EAI Endorsed Transactions on Cognitive Communications. He is a Reviewer for the IEEE Transactions on Vehicular Technology and other journals.
\end{biography}

\begin{biography}[{\includegraphics[width=1in,height=1.25in,clip,keepaspectratio]{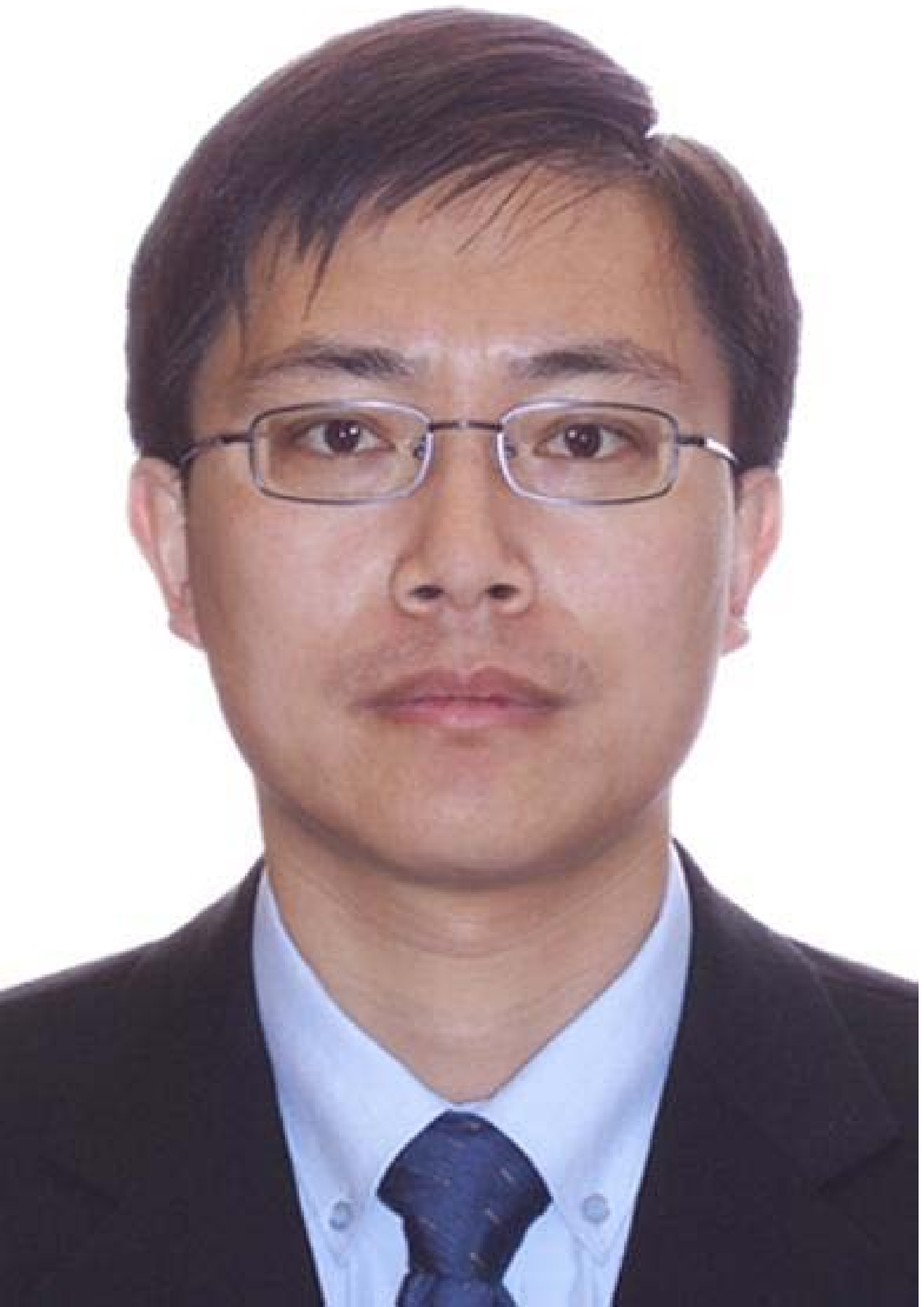}}]{Cheng-Xiang~Wang}
(S'01-M'05-SM'08) received the BSc and MEng degrees in Communication and Information Systems from Shandong University, China, in 1997 and 2000, respectively, and the PhD degree in Wireless Communications from Aalborg University, Denmark, in 2004.

He has been with Heriot-Watt University, Edinburgh, UK, since 2005, first as a Lecturer, then as a Reader in 2009, and was promoted to a Professor in 2011. He is also an Honorary Fellow of the University of Edinburgh, UK, and a Chair/Guest Professor of Shandong University and Southeast University, China. He was a Research Fellow at the University of Agder, Grimstad, Norway, from 2001--2005, a Visiting Researcher at Siemens AG-Mobile Phones, Munich, Germany, in 2004, and a Research Assistant at Technical University of Hamburg-Harburg, Hamburg, Germany, from 2000--2001. His current research interests include wireless channel modeling and simulation, green communications, cognitive radio networks, vehicular communication networks, massive MIMO, millimetre wave communications, and 5G wireless communication networks. He has edited 1 book and published 1 book chapter and over 190 papers in refereed journals and conference proceedings.

Prof. Wang served or is currently serving as an editor for 8 international journals, including \textit{IEEE Transactions on Vehicular Technology} (2011--) and \textit{IEEE Transactions on Wireless Communications} (2007--2009). He was the leading Guest Editor for \textit{IEEE Journal on Selected Areas in Communications, Special Issue on Vehicular Communications and Networks}. He served or is serving as a TPC member, TPC Chair, and General Chair for over 70 international conferences. He received the Best Paper Awards from IEEE Globecom 2010, IEEE ICCT 2011, ISIT 2012, and IEEE VTC 2013-Fall. He is a Fellow of the IET, a Fellow of the HEA, and a member of EPSRC Peer Review College. \end{biography}


\begin{thebibliography}{1}

\bibitem{Ge11}
    I. Humar, X. Ge, X. Lin, M. Jo and M. Chen, ``Rethinking energyefficiency models of cellular networks with embodied energy," \emph{IEEE Netw. Mag.}, vol. 25, no. 2, pp. 40-49, March 2011.

\bibitem{Li11}
    C.-X. Wang, F. Haider, X. Gao, X.-H. You, Y. Yang, D. Yuan, H. Aggoune, H. Haas, S. Fletcher, and E. Hepsaydir, ``Cellular architecture and key technologies for 5G wireless communication networks," \emph{IEEE Commun. Mag.}, vol. 52, no. 2, pp. 122-130, Feb. 2014.

\bibitem{Raghavendra12}
     S. Raghavendra, B. Daneshrad, ``Performance analysis of energy-efficient power allocation for MIMO-MRC systems," \emph{IEEE Trans. Commun.}, vol. 60, no. 8, pp. 2048--2053, Aug. 2012.

\bibitem{Liu08}
    J. Liu, Y. T. Hou, Y. Shi and D. S. Hanif, ``Cross-layer optimization for MIMO-based wireless ad hoc networks: routing, power allocation, and bandwidth allocation," \emph{IEEE J. Sel. Areas Commun.}, vol. 26, no. 6, pp. 913--926, Aug. 2008.

\bibitem{Ding10}
    J. Ding, D. Deng, T. Wu and H. Chen, ``Quality-aware bandwidth allocation for scalable on-demand streaming in wireless networks," \emph{IEEE J. Sel. Areas Commun.}, vol. 28, no. 3, pp. 366--376, Apr. 2010.

\bibitem{Su10}
    X. Su, S. Chan and J. H. Manton, ``Bandwidth allocation in wireless ad hoc networks: challenges and prospects," \emph{IEEE Commun. Mag.}, vol. 48, no. 1, pp. 80--85, Jan. 2010.

\bibitem{Helonde11}
    D. Helonde, V. Wadhai, V. S. Deshpande and H. S. Ohal, ``Performance analysis of hybrid channel allocation scheme for mobile cellular network," in \emph{Proc. IEEE ICRTIT 2011}, pp. 245--250, June 2011.

\bibitem{chengxiang07}
    C.-X. Wang, M. Patzold, and D. Yuan, ``Accurate and efficient simulation of multiple uncorrelated Rayleigh fading waveforms," \emph{IEEE Trans. Wireless Commun.}, vol. 6, no. 3, pp. 833-839, March 2007.

\bibitem{Xiang13}
    L. Xiang, X. Ge, C-X. Wang, F. Li, and F. Reichert, ``Energy Efficiency Evaluation of Cellular Networks Based on Spatial Distributions of Traffic Load and Power Consumption," \emph{IEEE Trans. Wireless Commun.}, vol. 12, no. 3, pp. 961--973, March 2013.

\bibitem{Chen11}
    C. Chen, W. Stark and S. Chen, ``Energy-bandwidth efficiency tradeoff in MIMO multi-hop wireless networks," \emph{IEEE J. Sel. Areas Commun.}, vol. 29, no. 8, pp. 1537--1546, Sep. 2011.

\bibitem{Heliot12}
    F. Heliot, M. A. Imran and R. Tafazolli, ``On the energy efficiency-spectral efficiency trade-off over the MIMO Rayleigh fading channel," \emph{IEEE Trans. Commun.}, vol. 60, no. 5, pp. 1345--1356, May 2012.

\bibitem{Wang13} I. Ku,  C. Wang, and J. S. Thompson, ``Spectral-energy efficiency tradeoff in relay-aided cellular networks��, \emph{IEEE  Trans. Wireless Commun.}, vol. 12, no. 10, pp. 4970-4982, Oct. 2013.

\bibitem{Hong13} X. Hong, Y. Jie, C. Wang, J. Shi, and X. Ge, ``Energy-spectral efficiency trade-off in virtual MIMO cellular systems,��  \emph{IEEE J. Sel. Areas Commun.}, vol. 31, no. 10, pp. 2128-2140, Oct. 2013.

\bibitem{Ku13}I. Ku,  C. Wang, and J. S. Thompson, ``Spectral, energy and economic efficiency of relay-aided cellular networks,�� \emph{IET Commun.}, vol. 7, no. 14, pp. 1476-1487, Sept. 2013.

\bibitem{Fisher15}
    R. A. Fisher, ``Frequency distribution of the values of the correlation coefficient in samples from an indefinitely large population," \emph{Biometrika}, vol. 10, pp. 507--521, 1915.

\bibitem{Wishart28}
    J. Wishart, ``The generalized product moment distribution in samples from a normal multivariate population," \emph{Biometrika}, vol. 20A, pp. 32--52, 1928.

\bibitem{Wishart48}
    J. Wishart, ``Proofs of the distribution law of the second order moment statistics," \emph{Biometrika}, vol. 35, pp. 55--57, 1948.

\bibitem{Matthaiou09}
    M. Matthaiou, D. I. Laurenson, and C.-X. Wang, ``On analytical derivations of the condition number distributions of dual non-central Wishart matrices," \emph{IEEE Trans. Wireless Commun.}, vol. 8, no. 3, pp. 1212-1217, Mar. 2009.

\bibitem{Zanella08}
    A. Zanella, M. Chiani and M. Z. Win, ``A general framework for the distribution of the eigenvalues of Wishart matrices," in \emph{Proc. IEEE ICC 2008}, pp. 1271--1276, May 2008.

\bibitem{Jin06}
    S. Jin, X. Gao and R. M. Matthew, ``Ordered eigenvalues of complex noncentral Wishart matrices and performance analysis of SVD MIMO systems," in \emph{Proc. IEEE ISIT 2006}, pp. 1564-1568, July 2006.

\bibitem{Zanella051}
    A. Zanella, M. Chiani and M. Z. Win, ``MMSE reception and successive interference cancellation for MIMO systems with high spectral efficiency," \emph{IEEE Trans. Wireless Commun.}, vol. 4, no. 3, pp. 1244--1253, May 2005.

\bibitem{Zanella052}
    A. Zanella, M. Chiani and M. Z. Win, ``Performance of MIMO MRC in correlated Rayleigh fading environments," in \emph{Proc. IEEE VTC 2005-Spring}, pp. 1633--1637, May 2005.

\bibitem{McKay07}
    M. R. McKay, A. J. Grant and I. B. Collings, ``Performance analysis of MIMO-MRC in double-correlated Rayleigh environments," \emph{IEEE Trans. Commun.}, vol. 55, no. 3, pp. 497--507, Mar. 2007.

\bibitem{Kang03}
    M. Kang, M. S. Alouini, ``Largest eigenvalue of complex Wishart matrices and performance analysis of MIMO MRC systems," \emph{IEEE J. Sel. Areas Commun.}, vol. 21, no. 3, pp. 418--426, Apr. 2003.

\bibitem{Kang04}
    M. Kang, M. S. Alouini, ``A comparative study on the performance of MIMO MRC systems with and without cochannel interference," \emph{IEEE Trans. Commun.}, vol. 52, no. 8, pp. 1417--1425, Aug. 2004.

\bibitem{Park06}
    C. S. Park, K. B. Lee, ``Statistical transmit antenna subset selection for limited feedback MIMO systems," in \emph{Proc. IEEE APCC 2006}, pp. 1--5, Aug. 2006.

\bibitem{Niyato10}
    D. Niyato, E. Hossain and K. Dong, ``Jiont admission control and antenna assignment for multiclass QoS in spatial multiplexing MIMO wireless networks," \emph{IEEE Commun. Mag.}, vol. 8, no. 9, pp. 4855--4865, Sep. 2010.

\bibitem{Karray10}
    M. K. Karray, ``Analytical evalution of QoS in the downlink of OFDMA wireless cellular networks serving streaming and elastic traffic," \emph{IEEE Trans. Commun.}, vol. 9, no. 5, pp. 1799--1807, May 2010.

\bibitem{Wu03}
    D. Wu, R. Negi, ``Effective capacity: a wireless link model for support of quality of  service," \emph{IEEE Trans. Wireless Commun.}, vol. 2, no. 4, pp. 630--643, July 2003.

\bibitem{Gursoy09}
    M. C. Gursoy, D. Qiao and S. Velipasalar, ``Analysis of energy efficiency in fading channels under QoS constraints," \emph{IEEE Trans. Wireless Commun.}, vol. 8, no. 8, pp. 4252--4263, Aug. 2009.


\bibitem{Tang071}
    J. Tang, X. Zhang, ``Quality-of-service driven power and rate adaptation over wireless links," \emph{IEEE Trans. Wireless Commun.}, vol. 6, no. 8, pp. 3058--3068, Aug. 2007.

\bibitem{Tang072}
    J. Tang, X. Zhang, ``Quality-of-service driven power and rate adaptation for multichannel communications over wireless links," \emph{IEEE Trans. Wireless Commun.}, vol. 6, no. 12, pp. 4349--4360, Dec. 2007.

\bibitem{Bogucka11}
    H. Bogucka, A. Conti, ``Degrees of freedom for energy savings in practical adaptive wireless systems," \emph{IEEE Commun. Mag.}, vol. 49, no. 6, pp. 38--45, June 2011.

\bibitem{Chiani03}
    M. Chiani, M. Z. Win and A. Zanella, ``On the capacity of spatially correlated MIMO Rayleigh fading channels," \emph{IEEE Trans. Inf. Theory}, vol. 49, no. 10, pp. 2363--2371, Oct. 2003.

\bibitem{Telatar99}
    E. Telatar, ``Capacity of multi-antenna Gaussian channels," \emph{Europ. Trans. Telecomm.}, vol. 10, pp. 585--595, Nov. 1999.

\bibitem{Kang06}
    M. Kang, M. S. Alouini, ``Capacity of MIMO Rician channels," \emph{IEEE Trans. Wireless Commun.}, vol. 5, no. 1, pp. 112--122, Jan. 2006.

\bibitem{Edelman89}
    A. Edelman, ``Eigenvalues and condition numbers of random matrices," Ph.D. dissertation, MIT, Cambridge, MA, May 1989.

\bibitem{Zhihua08}
    Z. Zhihua, X. He, and W. Jianhua, ``Average power control algorithmwith dynamic channel assignment for TDD-CDMA systems," in \emph{Proc. IEEE ICAIT 2008}, July 2008.

\end{thebibliography}
\end{document}